\pdfoutput=1 
\documentclass[times]{cnmauth}

\usepackage{moreverb}

\usepackage[colorlinks,bookmarksopen,bookmarksnumbered,citecolor=red,urlcolor=red]{hyperref}

\usepackage{url,lineno,microtype,subcaption}
\usepackage[onehalfspacing]{setspace}
\usepackage{color} 

\usepackage{import}
\usepackage{epsfig}

\usepackage{paralist}

\usepackage{amssymb}
\usepackage{amsmath}
\usepackage{bm}

\usepackage[mathscr]{euscript}
\usepackage{comment}

\usepackage[noabbrev,capitalise,english]{cleveref} 

\usepackage{framed} 
\usepackage{multicol} 

\usepackage{multirow} 

\usepackage[noprefix]{nomencl} 
\makenomenclature
\graphicspath{{./figs/}}

\usepackage{xcolor}
\colorlet{lightred}{red!70!white}

\usepackage[noadjust]{cite}

\usepackage{blindtext}

\renewcommand{\b}{\mathbf}

\newcommand{\eg}{e.g.} 
\newcommand{\ie}{i.e.} 

\makeatletter
\newcommand*{\rom}[1]{\expandafter\@slowromancap\romannumeral #1@}
\makeatother

\usepackage{lineno}

\newcommand\BibTeX{{\rmfamily B\kern-.05em \textsc{i\kern-.025em b}\kern-.08em
T\kern-.1667em\lower.7ex\hbox{E}\kern-.125emX}}

\def\volumeyear{2017}

\begin{document}

\runningheads{H.~P.~Bui et al.}{Corotational Cut Finite Element Method}


\title{Corotational Cut Finite Element Method for real-time surgical simulation: application to needle insertion simulation}

\author{Huu Phuoc Bui\corrauth \,$^{1,2}$, Satyendra Tomar\,$^{1}$ and St\'{e}phane P.A. Bordas\corrauth \,$^{1,3,4}$}

\address{
\begin{center}
$^{1}$Institute of Computational Engineering, University of Luxembourg, Faculty of Sciences Communication and Technology, Luxembourg \\
$^{2}$Laboratoire de Math\'{e}matiques de Besan\c{c}on (LMB) UMR CNRS 6623, Universit\'{e} de Franche-Comt\'{e}, Besan\c{c}on, France \\
$^{3}$Institute of Mechanics and Advanced Materials, School of Engineering, Cardiff University, UK \\
$^{4}$Intelligent Systems for Medicine Laboratory, University of Western Australia, Perth, Australia
\end{center}
}

\corraddr{huu-phuoc.bui@alumni.unistra.fr (H. P. Bui), stephane.bordas@alum.northwestern.edu (S. P. A. Bordas)}

\begin{abstract}
This paper describes the use of the corotational cut Finite Element Method (FEM) for real-time surgical simulation. Users only need to provide a background mesh which is not necessarily conforming to the boundaries/interfaces of the simulated object. The details of the surface, which can be directly obtained from binary images, are taken into account by a multilevel embedding algorithm applied to elements of the background mesh that cut by the surface. Boundary conditions can be implicitly imposed on the surface using Lagrange multipliers. The implementation is verified by convergence studies with optimal rates. The algorithm is applied to various needle insertion simulations (e.g. for biopsy or brachytherapy) into brain and liver to verify the reliability of method, and numerical results show that the present method can make the discretisation independent from geometric description, and can avoid the complexity of mesh generation of complex geometries while retaining the accuracy of the standard FEM. Using the proposed approach is very suitable for real-time and patient specific simulations as it improves the simulation accuracy by taking into account automatically and properly the simulated geometry. 
\end{abstract}


\maketitle


\section{Introduction}
\label{sec:Introduction}

Nowadays real-time simulation plays an important role in different fields: from graphic animation \cite{Musse2001}, fracture of stiff materials \cite{Muller2001}, to surgical training and simulation \cite{Cotin1999,Cotin2000,Monserrat2001,Courtecuisse2014}. In the medical context, surgical simulations are not only useful for training, but also helpful for pre-operative planning, and intra-operative guidance. Surgical simulations have to take into account interactions between a surgeon or an interventional radiologist with a deformable organ via surgical instruments (\eg{} a needle), and also interactions between the organ with its neighbouring structures. To be helpful, it is required that that computations to be performed in real time. To achieve real-time performance, some advanced solvers (\eg{} GPU-based computation \cite{COURTECUISSE2010}, or asynchronous solver \cite{Courtecuisse2014}) can be used. Coarse meshes can also be employed to reduce computational time. However, using coarse meshes, one may lose some geometric details, and simulations using coarse meshes are only suitable for targeted surgical training which relies more on visual realism than exact, but not for surgical planning or guidance where computations must provide accurate results. Model order reduction technique is also used to solve system equations with lower accuracy, but with significantly less time \cite{Niroomandi2012}. To reduce the computational effort, Quesada \emph{et al} \cite{Quesada2016} propose a computational parametric meta-model which is computed offline, and is only evaluated online.

Medical simulations have to deal with complex anatomical structures, \eg{} prostate, blood vessel, liver, brain, brain ventricle, etc. When the patient-specific geometry is considered, the mesh of the organ need to be reconstructed since the organ geometry is different from a patient to another. Misra and coworkers \cite{MISRA2009} have shown that the geometry of the organ and boundary conditions surrounding the organ are the most important factors influencing the organ deformation, and thus have a direct impact on the accuracy of simulation and planning. And when supercomputers or parallel computation is not considered, in order to response to real time simulation requirements, computations involving interactions with surgical tools and/or cutting operations, are performed on coarse meshes while applying pre-computed deformation from fine meshes, as proposed in \cite{Cotin2000}. However, these coarse and fine meshes may not conform to each other, resulting in different geometric description and different boundary conditions, and thus the question of simulation accuracy should be considered.



The use of geometric description in a computational method so that users only need to provide a background mesh which is not necessarily conforming to the boundary geometry of the simulated object, can dramatically reduce computational cost of preprocessing. This is the idea behind the fictitious domain method \cite{Burman2010,burman2012fictitious}, or the cut finite element method (CutFEM) \cite{Burman2015}. The extended finite element method (XFEM) \cite{Belytschko1999,Moes1999}, and the generalised finite element method (GFEM) \cite{Strouboulis2000} have been developed to deal with crack surfaces, or material interfaces evolving during simulation. These approaches allow for computation of coupled physical processes on distinct subregions of the total volume, and does not require an absolute conformity between the meshes from mesh construction as in \cite{Tabor2007}. In the context of real-time patient-specific surgical simulations, using an unified geometric description in a finite element code can not only importantly reduce preprocessing cost of mesh generation, but also can improve accuracy of simulation and planning.

The contribution of our paper is firstly on implementation aspect of the CutFEM using the so-called multilevel embedding approach to correctly integrate implicit boundaries of the simulated organ, and to accurately capture implicit interfaces of \eg{}, a tumour. The approach is implemented with a corotational model \cite{MOITA1996} which is suitably and widely used for the treatment of large rotations of soft tissues, see \eg{}, \cite{COURTECUISSE2010,Suwelack2012,Haouchine2013}. The implementation is verified by a convergence study revealing an optimal rate. We use Lagrange multipliers to implicitly impose Dirichlet and Neumann boundary conditions on implicit boundaries. We demonstrate the performance of the corotational CutFEM through various applications from needle insertion simulations (\eg{} for biopsy or brachytherapy) to simulation of electrode lead implantation in Deep Brain Stimulation (DBS) procedure. The algorithm is implemented in open-source SOFA framework \cite{Faure2012_SOFA} \footnote{https://www.sofa-framework.org}.

The remaining of the paper is organised as follows. In \cref{sec:Methods}, we describe the formulation of the needle insertion problem into soft tissues, together with its the discrete form. We present geometry discretisation of cut elements, and the algorithm for multilevel embedding approach to correctly integrate implicit boundaries/interfaces. The implementation aspect is discussed as well. Then, a corotational formulation for CutFEM is shortly introduced. It is followed by the description of how boundary conditions are applied on implicit surfaces. We also discuss on solving the system equations with constraints. Numerical results are presented in \cref{sec:Results}, which demonstrate the capabilities of the approach through various needle insertion problems into the liver, and into the brain. And finally, conclusions are drawn in \cref{sec:Conclusions}.


\section{Methods}
\label{sec:Methods}

\subsection{Problem setting}

\begin{figure}[!htbp]
 \centering
       \begin{subfigure}[b]{0.5\textwidth}
	  \centering
	  \def\svgwidth{1\textwidth}
	  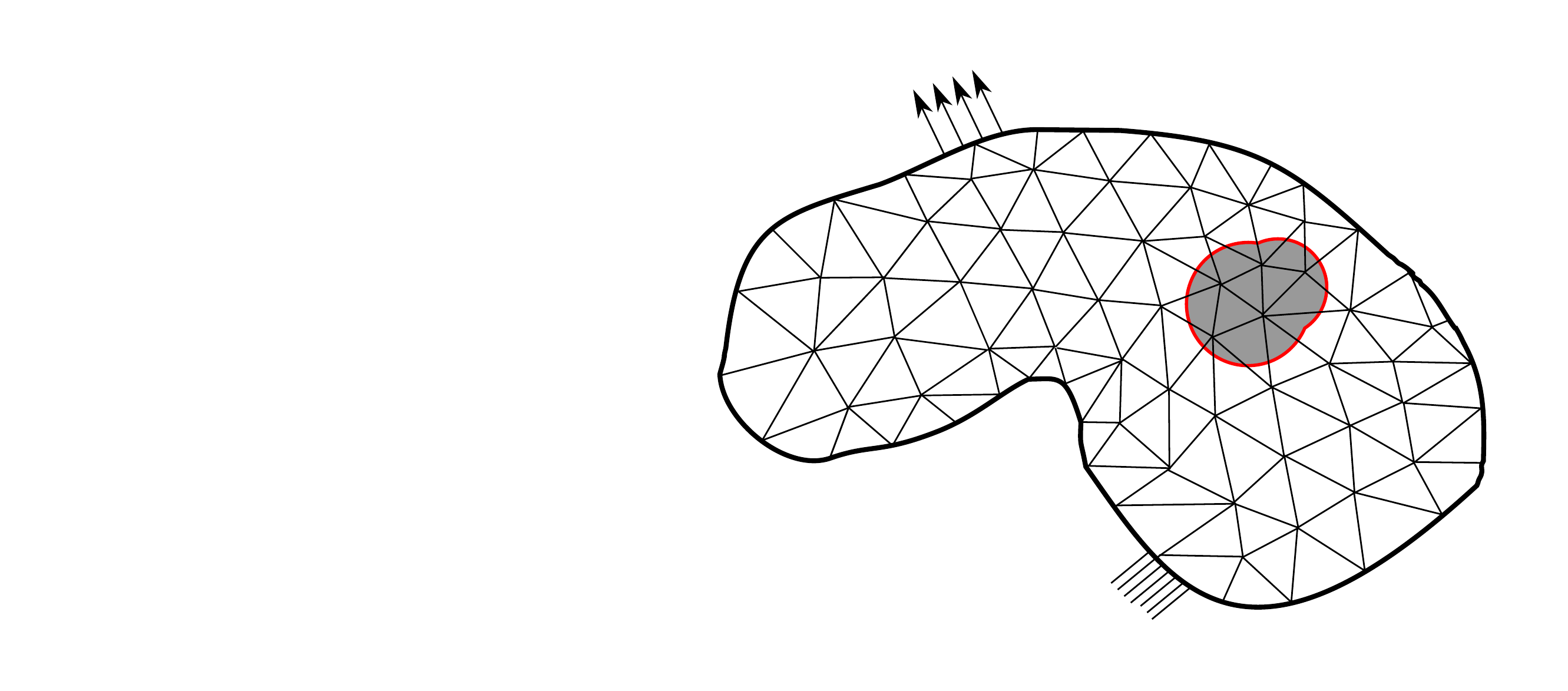
	  \caption{Immersed interface problem.}
	  \label{fig:immersed_intereface_problem}
      \end{subfigure}%
            ~ 
      \begin{subfigure}[b]{0.5\textwidth}
	  \centering
	  \def\svgwidth{1\textwidth}
	  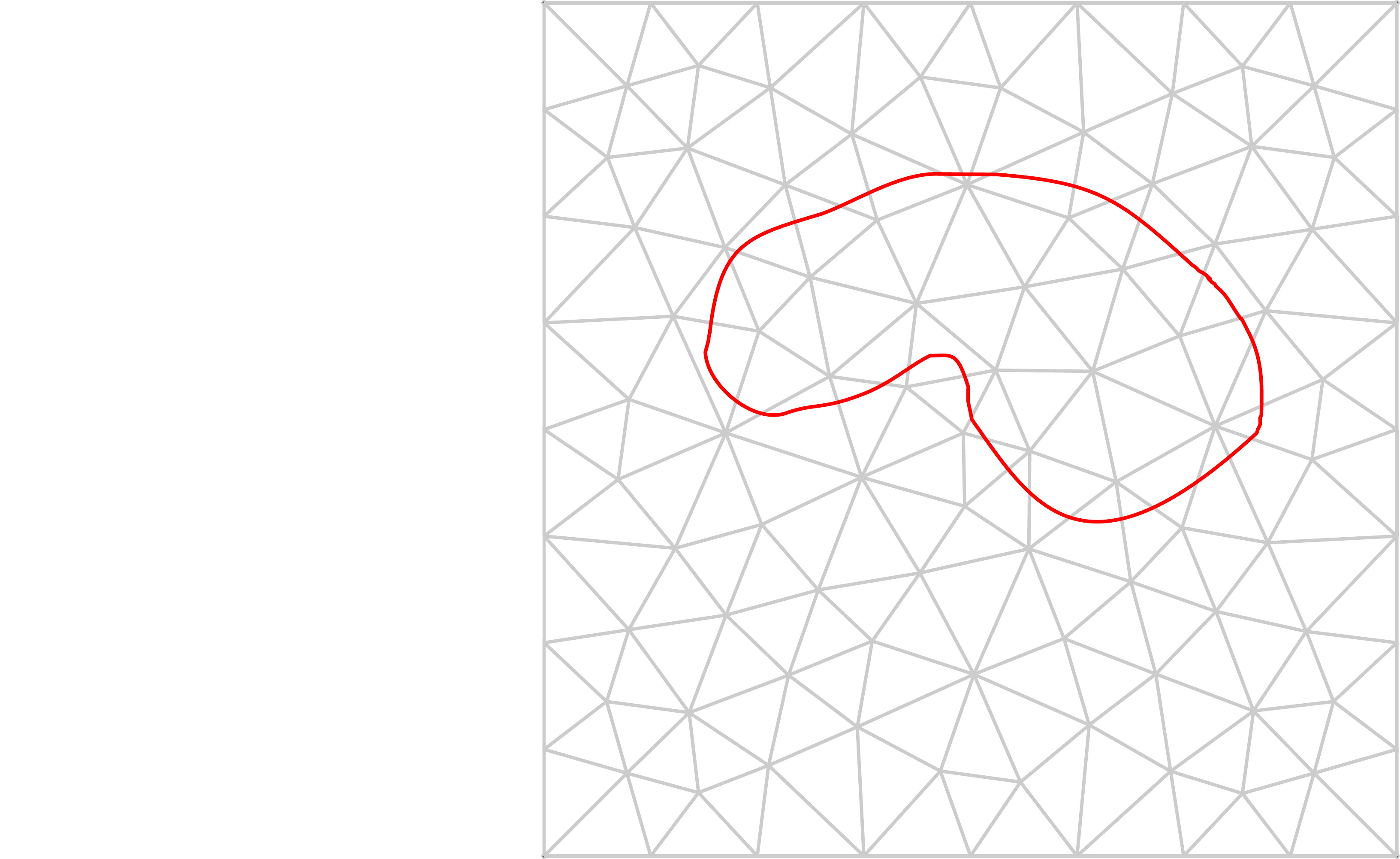
	  \caption{Implicit boundary problem.}
	  \label{fig:implicit_boundary_problem}
      \end{subfigure}%
\caption{Two dimensional representation of the studying problems.}\label{fig:implicite_interface_and_boundary_problem}
\end{figure}

In the context of needle insertion into soft tissue, we model both needle and tissue as dynamic deformable objects. In the case where the tissue is modelled as heterogeneous material with complex internal structures or the tissue has the complex geometry, using cut FEM is advantageous because it does not require conforming meshes. \cref{fig:immersed_intereface_problem} schematically shows a problem in which an interface is immersed into a tissue geometry for simulating \eg{} a tumour geometry. \cref{fig:implicit_boundary_problem} shows a problem in which the tissue is simulated with an implicit boundary (\ie{} the computational mesh is not fitted to the tissue geometry).

Let $\Omega$ and $\partial \Omega$ denote the domain and its boundary, respectively. The tissue undergoes an imposed displacement $\b{\bar{u}}$ on the boundary part $\Gamma_u$ and a traction force $\b{\bar{t}}$ on the boundary part $\Gamma_t$. The governing equation of the problem reads as
\begin{subequations}
\begin{align}
 \mathrm{div}\, \bm{\sigma} + \mathbf{b} + \bm{\lambda} & =  \rho \ddot{\mathbf{u}} \qquad \text{in } \Omega \label{eq:equilibrium}\\
 \bm{\epsilon} & =  \frac{1}{2}\left(\mathrm{grad}\,\mathbf{u} + (\mathrm{grad}\,\mathbf{u})^T \right)  \label{eq:kinematic} \\
 \bm{\sigma} & =  f(\bm{\epsilon},\bm{\nu}) \label{eq:constitutive}\\
 \bm{\sigma} \cdot \mathbf{n} & = \mathbf{\bar{t}} \; \text{on } \Gamma_t \label{eq:NeumannBC}; \\
 \mathbf{u}  &= \mathbf{\bar{u}} \; \text{on } \Gamma_u, \label{eq:DirichletBC}
\end{align}
\end{subequations}
where $\bm{\sigma}$ is the Cauchy stress tensor, $\mathbf{b}$ is the body force vector, $\rho$ is the mass density, $\bm{\epsilon}$ is the strain tensor, the material law is expressed through a relationship between the stress $\bm{\sigma}$ and the strain $\bm{\epsilon}$ via the vector of internal variables $\bm{\nu} = (\nu_1, \nu_2, \dots, \nu_n)$,  $\mathbf{u}$ is the displacement field of the object, and $\ddot{x}$ is the second partial derivative of $x$ with respect to time, $\mathbf{n}$ denotes the outward unit normal vector on $\Gamma_t$, and $\bm{\lambda}$ denotes the interaction force between the needle and the tissue.

The interaction force $\bm{\lambda}$ is defined from the interaction law between the needle and the tissue. Three type of constraints between the needle and the tissue are defined during needle insertion simulations: constraint puncture at tissue surface, constraint at needle tip, and constraints along the needle shaft, see \cite{Bui2016_TBME}.






\subsection{Weak form and FEM discretisation}
\label{sec:weakForm}

Since the partial differential equation \eqref{eq:equilibrium}, which states the equilibrium of the system, involves both spatial and temporal derivatives, it can be solved numerically by discretising that equation in both space (the volume representing the object) and time.

Using $N_n$ nodes, the domain $\Omega$, which represents tissue or needle, is spatially discretised into $N_e$ finite elements $\Omega_e$, $e=1,2,\dots,N_e$, see Figure~\ref{fig:implicite_interface_and_boundary_problem}. By integrating the equilibrium equation on each element volume, and assembling for the whole volume, we obtain the discrete problem as (see,~\eg{},~\cite{zienkiewicz2000finite,Liu201443})
\begin{equation}
\label{eq:weakForm}
 \mathbf{M} \ddot{ \mathbf{u} } + \mathbf{C} \dot{ \mathbf{u} } + \mathbf{K} \mathbf{u} = \mathbf{f}^{ext} + \mathbf{H}^T \bm{\lambda},
\end{equation}
where $\mathbf{M}$ is the mass matrix, $\mathbf{K}$ is the stiffness matrix, $\mathbf{C}$ is the damping matrix, and $\mathbf{f}^{ext}$ is the external force vector. The interaction force constraints $\bm{\lambda}$, between the needle and the tissue, are computed using Lagrange multipliers, $\mathbf{H}^T$ provides the direction of the constraints.
\cref{eq:weakForm} can be rewritten as
\begin{equation}
\label{eq:Ma=f}
 \mathbf{M} \mathbf{a} = \mathbf{f}(\mathbf{x},\mathbf{v}) + \mathbf{H}^T \bm{\lambda},
\end{equation}
where $\mathbf{a} = \ddot{\mathbf{u}}$, $\mathbf{x}$, $\mathbf{v} = \mathbf{\dot{\mathbf{u}}}$ are the acceleration, position and velocity vectors, respectively, and $\mathbf{f}(\mathbf{x},\mathbf{v}) = \mathbf{f}^{ext}-\mathbf{K}\mathbf{u} - \mathbf{C}\mathbf{v}$ represents the net force (the difference of the external and internal forces) applied to the object. 
%

%
For temporal discretisation, i.e. to numerically solve the problem in time, we use an implicit backward Euler scheme~\cite{Baraff1998}, which is described as follows
\begin{equation}
\label{eq:EulerBackward}
\dot{\mathbf{u}}_{t+\tau} = \dot{\mathbf{u}}_t + \tau\ddot{\mathbf{u}}_{t+\tau}; \quad \mathbf{u}_{t+\tau} = \mathbf{u}_{t} + \tau \dot{\mathbf{u}}_{t+\tau},
\end{equation}
where $\tau$ denotes the time step. Inserting \cref{eq:EulerBackward} into \cref{eq:Ma=f} yields the final discrete system
\begin{equation}
\label{eq:Ax=b}
\underbrace{(\mathbf{M}-\tau \mathbf{C}-\tau^2 \mathbf{K})}_{\mathbf{A}} d\mathbf{v} = \underbrace{ \tau\mathbf{f}(\mathbf{x}^t,\mathbf{v}^t) + \tau^2 \mathbf{K}\mathbf{v}^t}_{\mathbf{b}} + \mathbf{H}^T \bm{\lambda}
\end{equation}
or simply $\mathbf{A} d\mathbf{v} = \mathbf{b} + \mathbf{H}^T \bm{\lambda}$, where $d\mathbf{v} = \mathbf{v}_{t+\tau} - \mathbf{v}_t$.
After solving~\eqref{eq:Ax=b} for $d\mathbf{v}$, the position and velocity are updated for needle and tissue as
\begin{equation}
\label{eq:updateVX}
\mathbf{v}_{t+\tau} = d\mathbf{v} + \mathbf{v}_t; \quad \mathbf{x}_{t+\tau} = \mathbf{x}_{t} + \tau \mathbf{v}_{t+\tau}.
\end{equation}

The tissue domain is discretised by tetrahedron mesh. For the tetrahedra that intersect with the immersed interface/implicit boundary, an embedded element set is employed to facilitate the integration, see \cref{sec:discretisationGeometryForImplicitBoundaries}.

Since the needle geometry has a special character: its length is much greater than the dimensions of its cross section, a special assumption can be made during the needle deformation. It states that material points on the normal to the midline (\ie{} the neutral line) of the needle remain on the normal during deformation. With this, it makes possible to use a Euler-Bernoulli beam theory \cite{timoshenko1953history} to describe the behaviour of the needle. The needle is then discretised by one dimensional elements and one employs Hermite shape functions ($C^1$ continuous) for interpolation of the displacement field so that bending behaviour of the needle is well taken into account. By this, each node of the elements used for the needle has $6$ degrees of freedom ($3$ translations and $3$ rotations).

In this contribution, a lumped mass matrix, in which a diagonal mass matrix (from the mass density $\rho$) is integrated over the volume of each element is employed. The stiffness matrix $\mathbf{K}$ is computed based on the corotational FEM, see \cref{sec:corotationalFormulation}, for both the needle and tissue. 

\subsection{Discretising geometry for immersed/implicit boundaries}
\label{sec:discretisationGeometryForImplicitBoundaries}




The discretisation technique presented here is applicable for both fictitious domain problems \cite{Burman2010,Sotiropoulos2014} and nonconforming interface problems, see, \eg{} \cite{Qin2017}.

\subsubsection{Discretisation of geometry.}
\label{sec:discretisationGeometry}
\begin{figure}[!htbp]
 \centering
       \begin{subfigure}[b]{0.3\textwidth}
            \centering
            \def\svgwidth{\textwidth}
            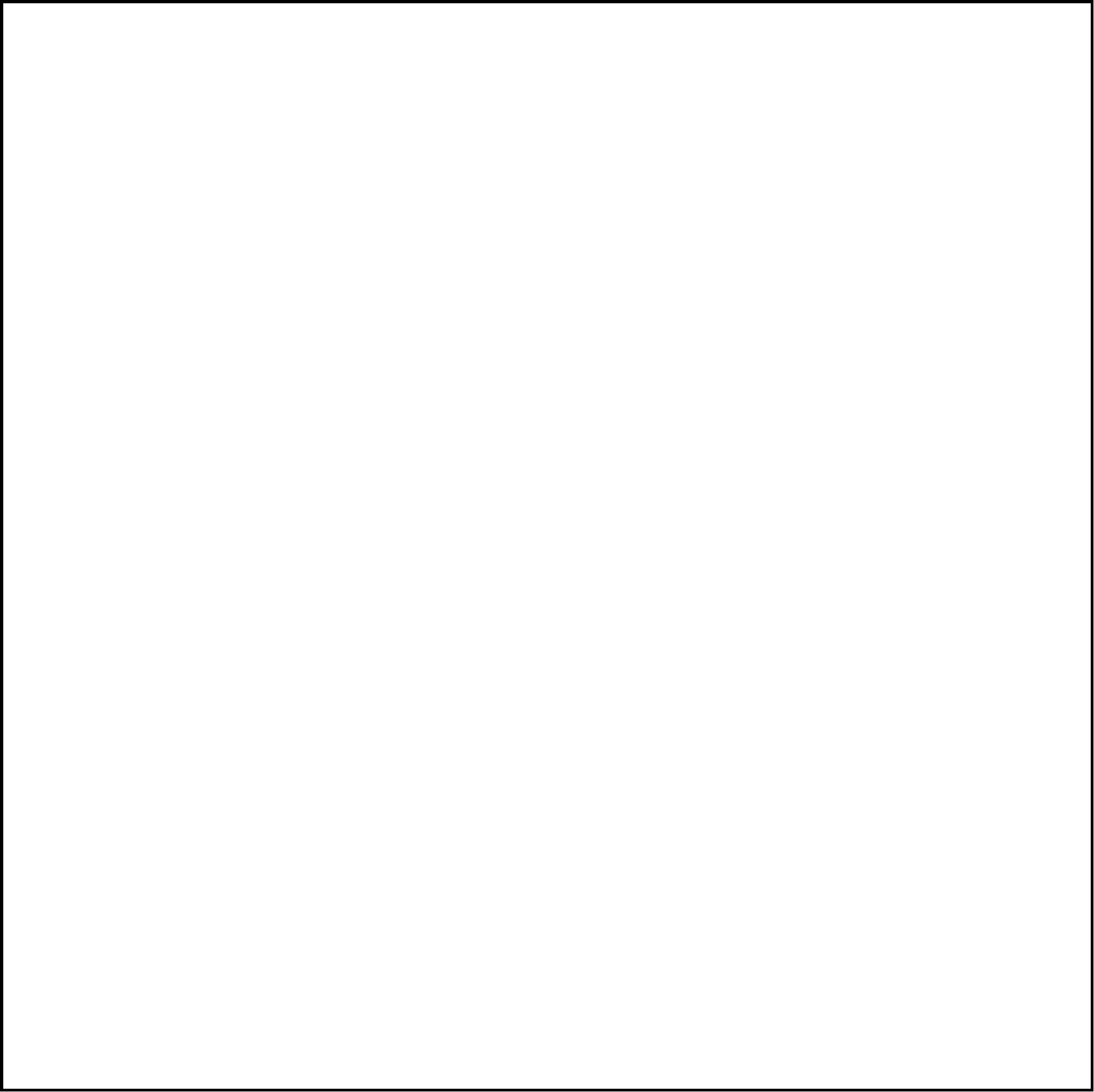
            \caption{}
            \label{fig:domain_definition}
      \end{subfigure}%
            ~ 
      \begin{subfigure}[b]{0.304\textwidth}
              \centering
	      \includegraphics[width=1\textwidth]{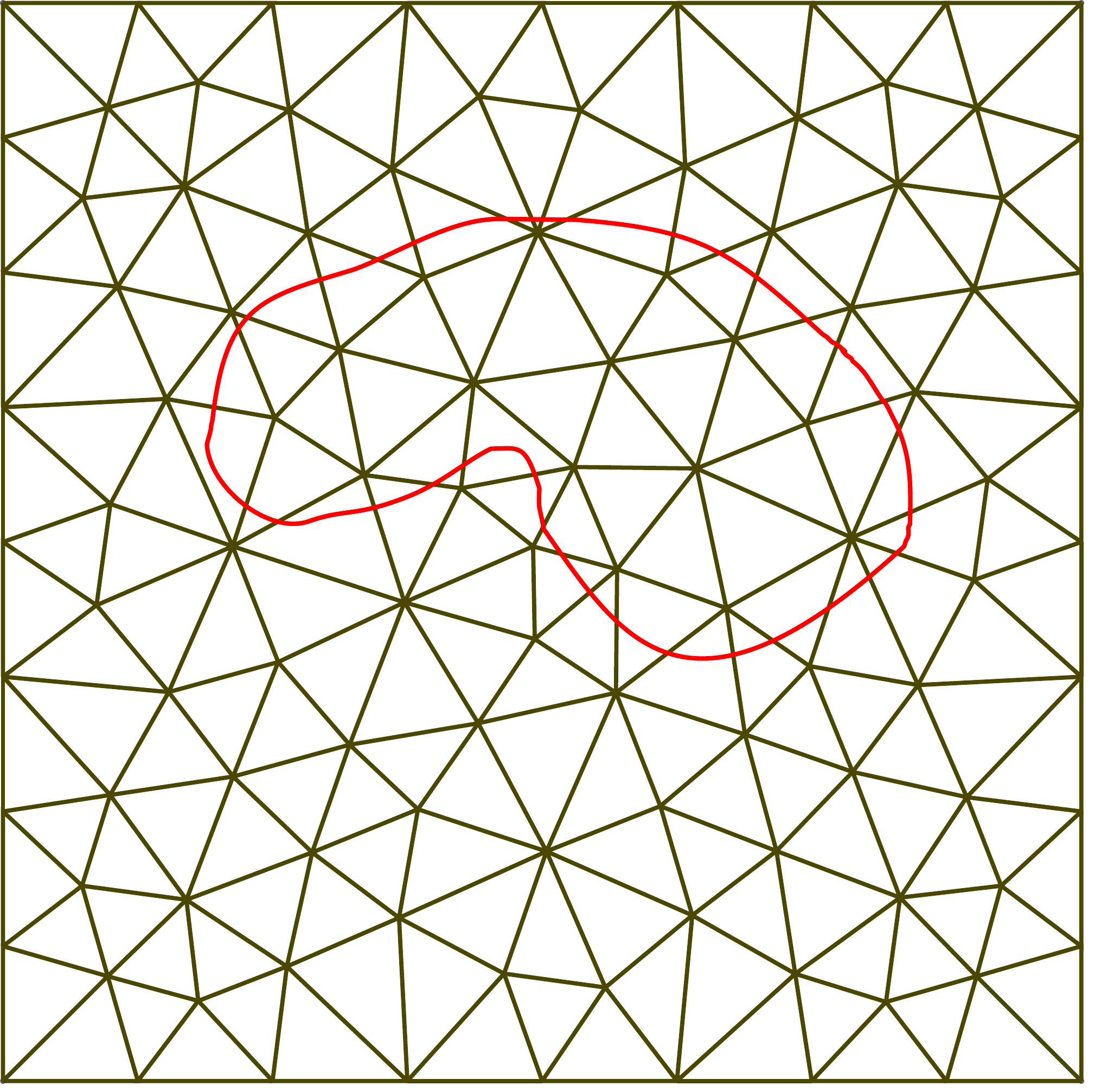}
              \caption{}
              \label{fig:meshing_domain}
      \end{subfigure}%
        ~ 
      \begin{subfigure}[b]{0.304\textwidth}
              \centering
	      \includegraphics[width=1\textwidth]{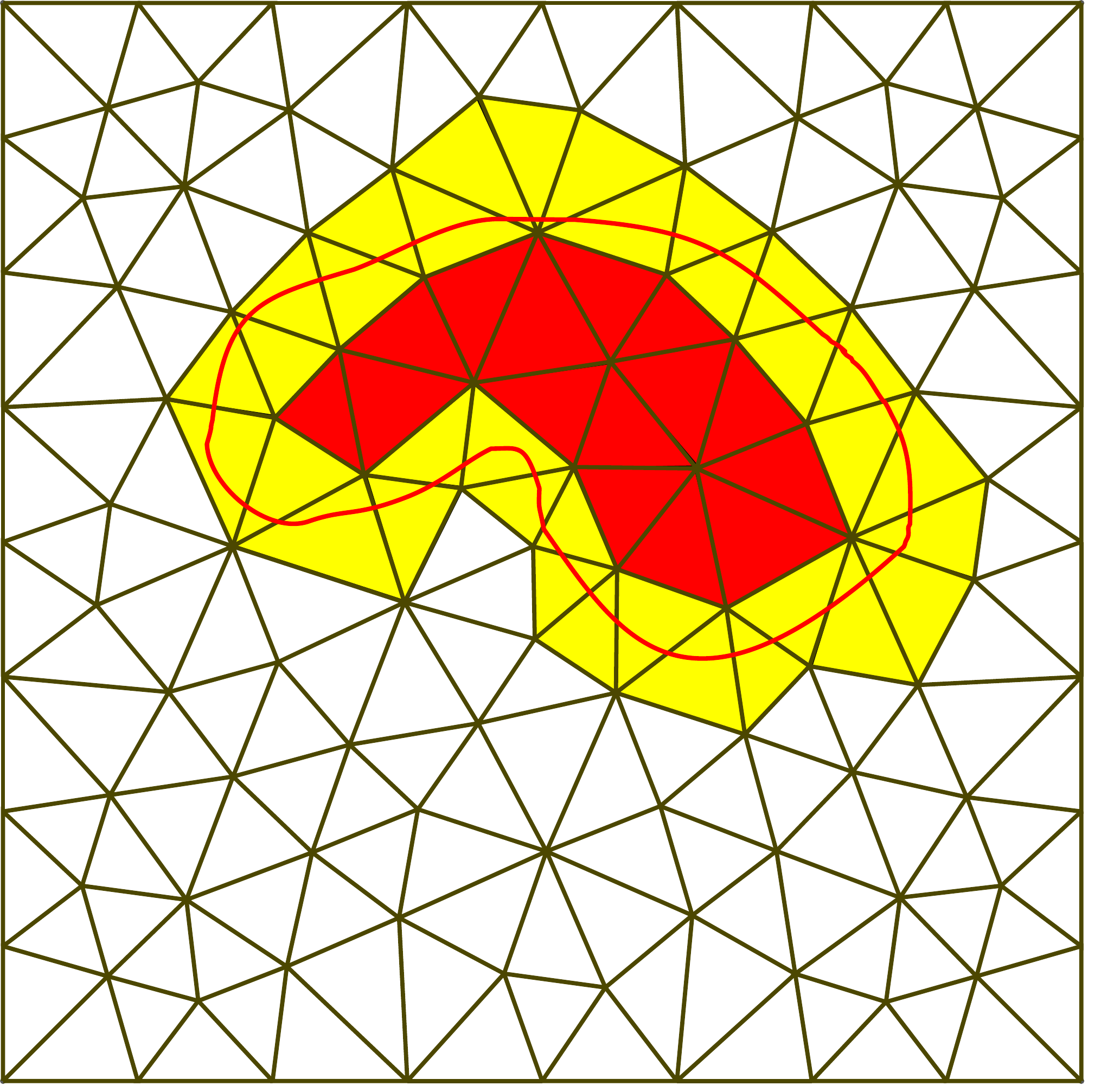}
              \caption{}
              \label{fig:meshing_domain_cutElementDetection}
      \end{subfigure}%
\caption{Two dimensional representation of the problem domains (\subref{fig:domain_definition}), the domains are nonconformingly discretised with respect to the surface $\Gamma$ (\subref{fig:meshing_domain}), the cut elements are highlighted by yellow colour whereas all inside elements of the volume $\Omega_2$ are highlighted by red colour, the remaining elements are marked as outside (\ie{} outside of $\Omega_2$) elements (\subref{fig:meshing_domain_cutElementDetection}).}\label{fig:domain_definition_and_meshing}
\end{figure}

In fictitious domain or nonconforming interface approaches, the boundary of a given domain is embedded into a computational mesh. The latter is used to approximate the solution of the governing PDEs. In general, the boundary surface of the given domain is represented by a very fine mesh. The boundary surface mesh and the computational mesh are not conforming. In fictitious domain method, one only integrates the governing equations on the inside volume bounded by the surface boundary (\ie{} $\Omega_2$), whereas in interface problems, during integration, different mechanical properties are assigned to the outside and inside volumes, $\Omega_1$ and $\Omega_2$, respectively separated by the surface, refer to \cref{fig:domain_definition}. The domain is discretised by a nonconforming mesh with respect to the surface $\Gamma$ as schematically shown in \cref{fig:meshing_domain}.

In order to integrate correctly on a given domain, the elements are firstly classified into three categories: cut elements, inside (\ie{} inside of $\Omega_2$) and outside (\ie{} outside of $\Omega_2$) elements, see \cref{fig:meshing_domain_cutElementDetection}. Then, each cut element is embedded with a sub-element set consisting of elements which are conforming with the surface $\Gamma$, to facilitate integration. It is noted that the sub-element set is only used for integration purpose, and thus it does not affect the approximation properties of the discretisation at all since the degrees of freedom of the cut elements are still only defined on their nodes.

\begin{figure}[!htbp]
  \centering
  \def\svgwidth{0.5\textwidth}
  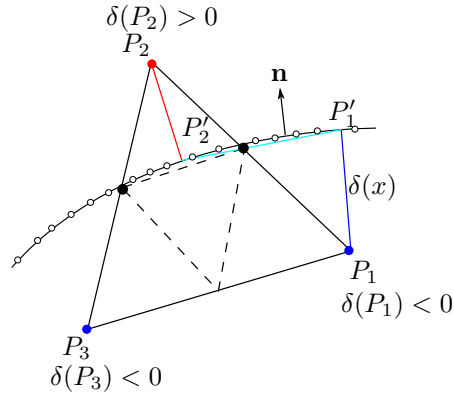
  \caption{Schematic representation of intersection computation between a triangle and a discretised surface. An embedded subtriangulation is performed on the element for integration purpose.}
  \label{fig:intersection_determination}
\end{figure}

To identify the cut elements, we use the level-set method \cite{Sethian1999level}. \cref{fig:intersection_determination} schematically shows an element which is potentially cut by the interface. To know if the element is really cut by the interface, we define the level-set function as a distance function from the nodes of the element to the interface and check the sign of that function. An edge $P_iP_j$ of the element is cut by the interface if and only if
\begin{equation}
\label{eq:conditionCut}
 \delta(P_i) \cdot \delta(P_j) < 0,
\end{equation}
where $P_i$ and $P_j$ are the position of two ends of the edge, $\delta$ is the distance defined from the points to the interface, see \cref{fig:intersection_determination}.

The intersection between the interface and an edge is approximated by the zero level-set. When \cref{eq:conditionCut} is fulfilled, the barycentric coordinate of the intersection reads
\begin{equation}
 \xi = \frac{ \lvert \delta(P_i) \rvert }{ \lvert \delta(P_i) \rvert + \lvert \delta{P_j} \rvert },
\end{equation}
and the position of the intersection is computed as
\begin{equation}
 P = (1-\xi) P_i + \xi P_j.
\end{equation}

It is noted that, the immersed interface in general applications has a complex geometry, and is not necessarily a convex surface, see \cref{fig:liver_in_tetrahedron_mesh_cropped}. Therefore, using the level set function to compute the inside and outside elements may not be efficient, especially when the elements are far from the interface. An efficient algorithm is presented in \cref{sec:implementation} to identify inside and outside elements when the cut elements are already marked using the level-set method.

Embedding each cut element by a conforming subtriangulation for integration purpose is detailed in \cref{sec:implementation}.

\subsubsection{Refinement for invalid cut elements.}
\label{sec:refinement_invalid_elements}

\begin{figure}[!htbp]
  \centering
  \includegraphics[width=1\columnwidth]{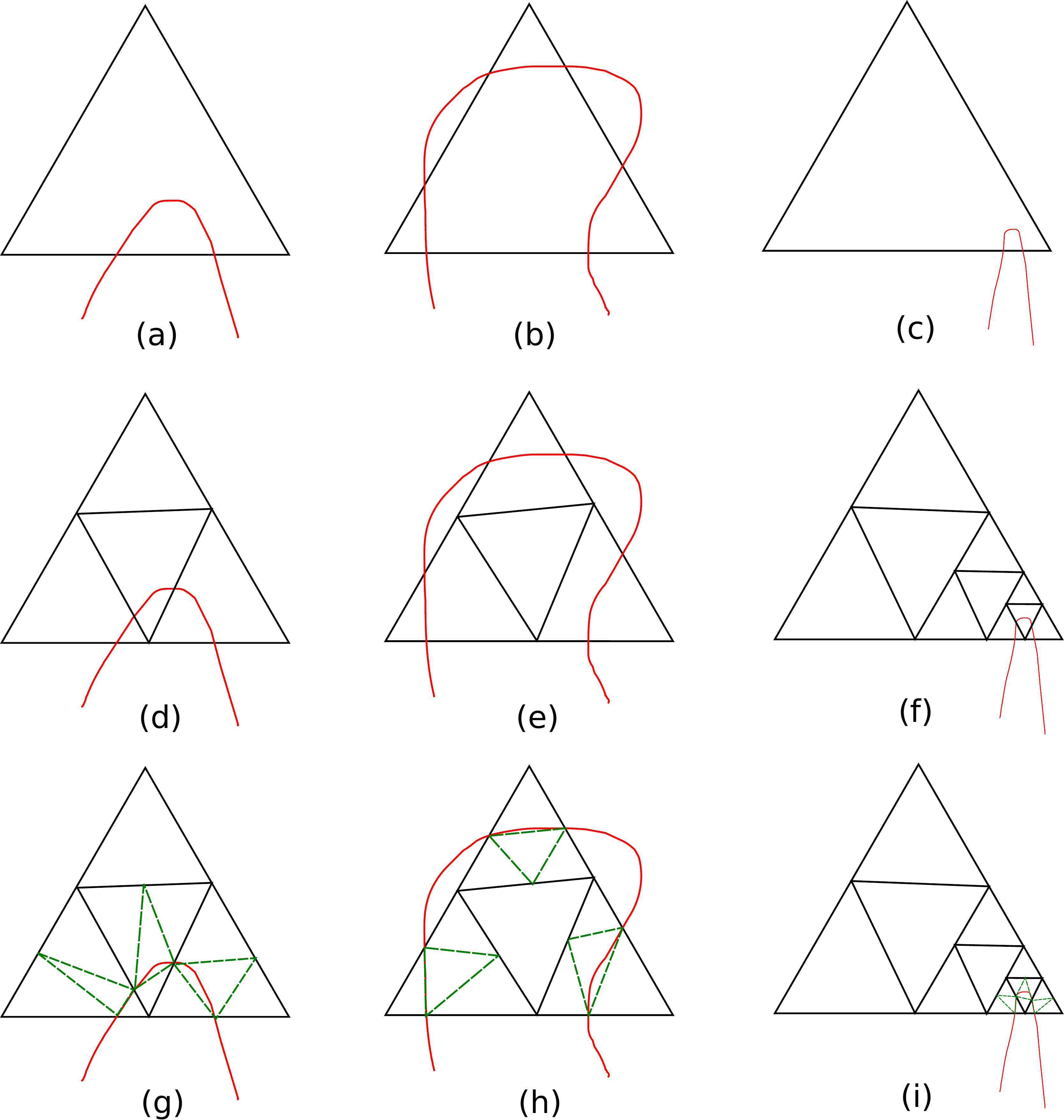}
  \caption{Schematic representation of invalid cut elements in two dimensions: (a), (b), (c); refinements lead to valid cut elements: (d), (e), (f). Valid cut elements are embedded with a subtetrahedron set (shown in green colour) for integration purpose: (g), (h), (i).}
  \label{fig:invalid_cut_elems}
\end{figure}

In \cref{sec:discretisationGeometry}, we assume that elements of the background mesh are cut by the surface interface with \emph{valid} cases: \ie{} the intersection between a tetrahedron element and the surface is either a triangle or a quadrilateral, see \cref{fig:tetrahedron_plane_intersection}. The number of intersection between the surface and the element edges is either 3 or 4. However, \emph{invalid} cutting cases arise when, for example, an edge of an element is cut by the surface with more than one intersection, or there are only two, or more than 4 edges of a tetrahedron, which are cut by the surface. These invalid cases can arrive when a coarse background mesh is used together with a \emph{curved} interface surface. Figures \ref{fig:invalid_cut_elems}a, \ref{fig:invalid_cut_elems}b, and \ref{fig:invalid_cut_elems}c schematically show, in two dimensions, some invalid cut cases between a triangle and an interface curve. 

To overcome this issue, one solution, as in \cite{Fries2016}, is to \emph{recursively} embed the invalid cut element with a set of subelements until all (sub) cut elements are valid, see Figures \ref{fig:invalid_cut_elems}d, \ref{fig:invalid_cut_elems}e, and \ref{fig:invalid_cut_elems}f. This procedure is called refinement in what follows. However, it is important to note that the invalid cut element is not actually refined since we do not introduce any new degree of freedom into the background mesh. The number of level of refinement needed to get all valid cut elements depends not only on the coarseness of the background mesh and curvature of the interface surface, but also on the relative location between the cut element and the surface. In Figures \ref{fig:invalid_cut_elems}d, \ref{fig:invalid_cut_elems}e, only one refinement level is needed in order to get valid cut elements, whereas in Figure \ref{fig:invalid_cut_elems}f, two refinement levels are necessary.

Once we get all valid cut elements, they are embedded by a conforming subtriangulation as usual for integration reason, described above. Figures \ref{fig:invalid_cut_elems}g, \ref{fig:invalid_cut_elems}h, and \ref{fig:invalid_cut_elems}i show the conforming embedded elements for valid cut (sub) elements.

\begin{figure}[!htbp]
 \centering
       \begin{subfigure}[b]{0.32\textwidth}
              \centering
	      \includegraphics[width=1\columnwidth]{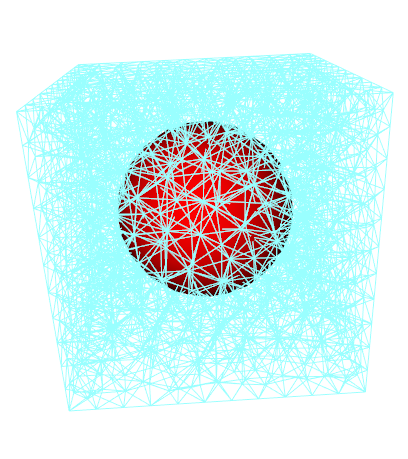}
              \caption{}
              \label{fig:sphere_immersed_in_background_mesh}
      \end{subfigure}%
             ~ 
      \begin{subfigure}[b]{0.32\textwidth}
              \centering
	      \includegraphics[width=0.7\columnwidth]{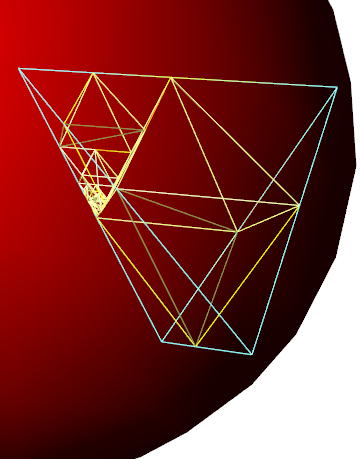}
	      \caption{}
	      \label{fig:invalid_cut_tetrahedron_refinements}
      \end{subfigure}%
      \begin{subfigure}[b]{0.32\textwidth}
              \centering
	      \includegraphics[width=1\columnwidth]{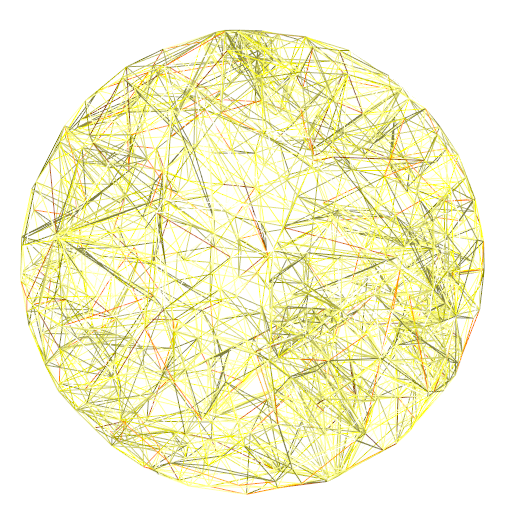}
              \caption{}
              \label{fig:sphere_multi_refinement_embedded_tetra}
      \end{subfigure}%
\caption{A sphere is immersed into a background mesh (\subref{fig:sphere_immersed_in_background_mesh}), an invalid cut element is embedded with subtetrahedra after refinement (\subref{fig:invalid_cut_tetrahedron_refinements}), and finally all, invalid and valid, cut elements, intersected with the spherical surface, are embedded with subtetrahedra, with or without refinement, respectively, for integration reason (\subref{fig:sphere_multi_refinement_embedded_tetra}).}\label{fig:refinement_for_sphere}
\end{figure}

As an example to demonstrate the implemented algorithm which works on tetrahedra, \cref{fig:sphere_immersed_in_background_mesh} shows a spherical surface which is immersed into a background mesh, and valid and invalid cut elements with the spherical surface are immersed with subtetrahedra shown in \cref{fig:sphere_multi_refinement_embedded_tetra}. \cref{fig:invalid_cut_tetrahedron_refinements} shows four levels of refinement needed to capture the intersections between an invalid cut tetrahedron and the spherical surface.


\subsubsection{Implementation aspects.}
\label{sec:implementation}





\begin{figure}[!htbp]
 \centering
       \begin{subfigure}[b]{0.4\textwidth}
              \centering
	      \includegraphics[width=1\columnwidth]{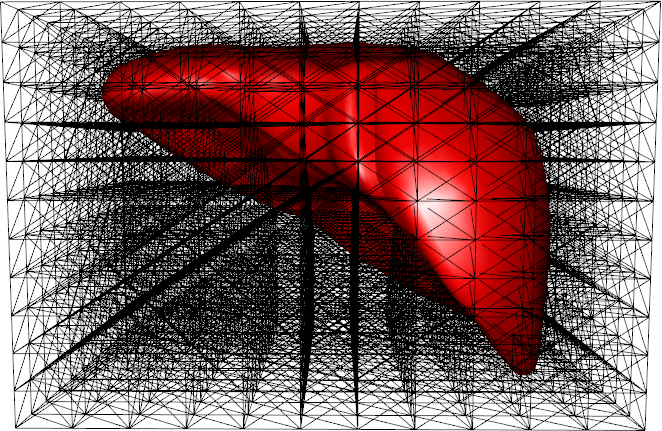}
              \caption{}
              \label{fig:liver_in_tetrahedron_mesh_cropped}
      \end{subfigure}%
             ~ 
      \begin{subfigure}[b]{0.45\textwidth}
              \centering
	      \includegraphics[width=1\columnwidth]{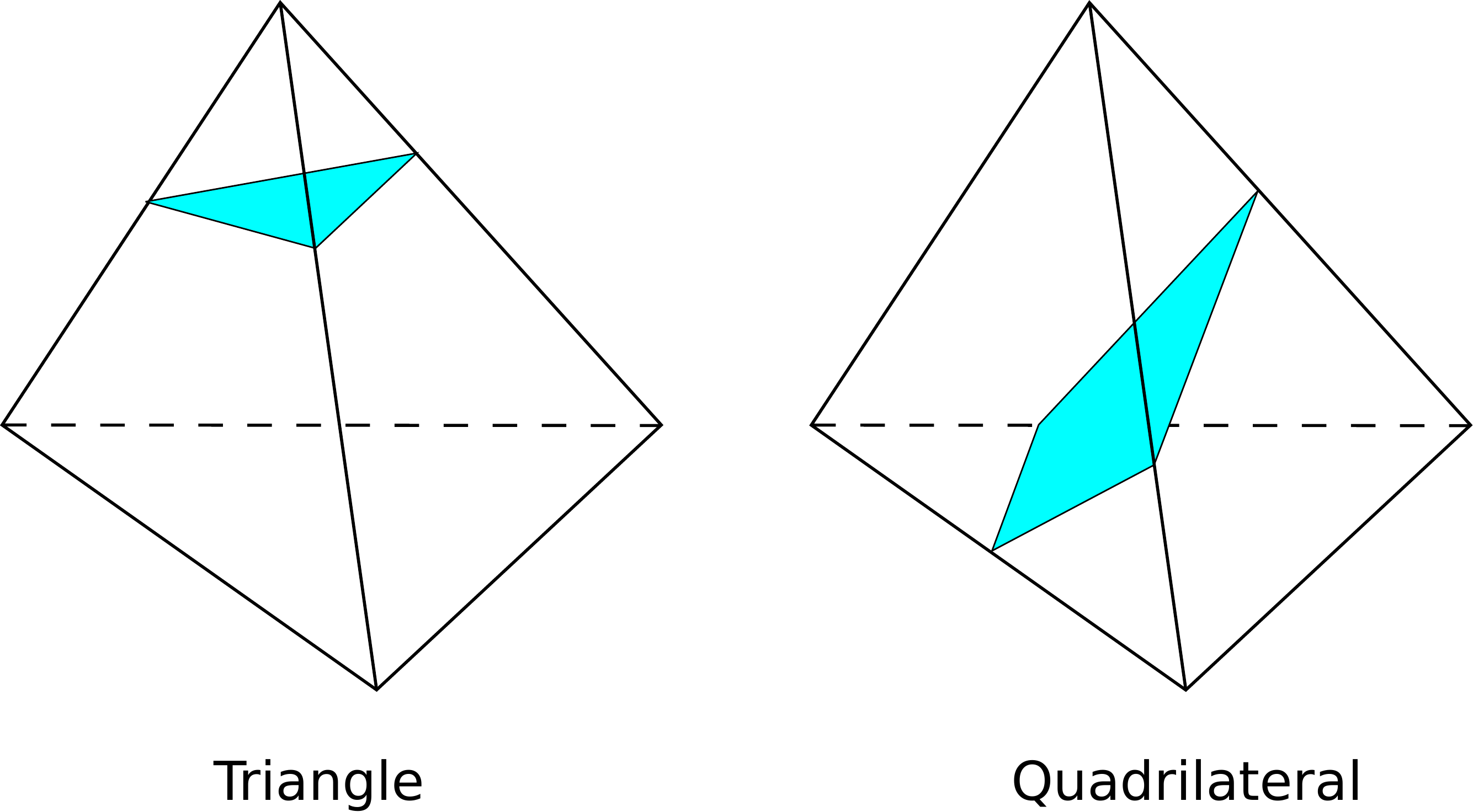}
              \caption{}
              \label{fig:tetrahedron_plane_intersection}
      \end{subfigure}%
\caption{A liver surface is immersed in a computational tetrahedron mesh (\subref{fig:liver_in_tetrahedron_mesh_cropped}), and intersection cases between a surface and a tetrahedron (\subref{fig:tetrahedron_plane_intersection}).}\label{fig:CutFEM_intersection}
\end{figure}

The starting point is that an arbitrary surface is immersed into a computational background mesh. As an example, a liver surface being immersed in a computational tetrahedron mesh is shown in \cref{fig:liver_in_tetrahedron_mesh_cropped}.

\begin{figure}[!htbp]
 \centering
       \begin{subfigure}[b]{0.6\textwidth}
	  \centering
	    \def\svgwidth{\textwidth}
	    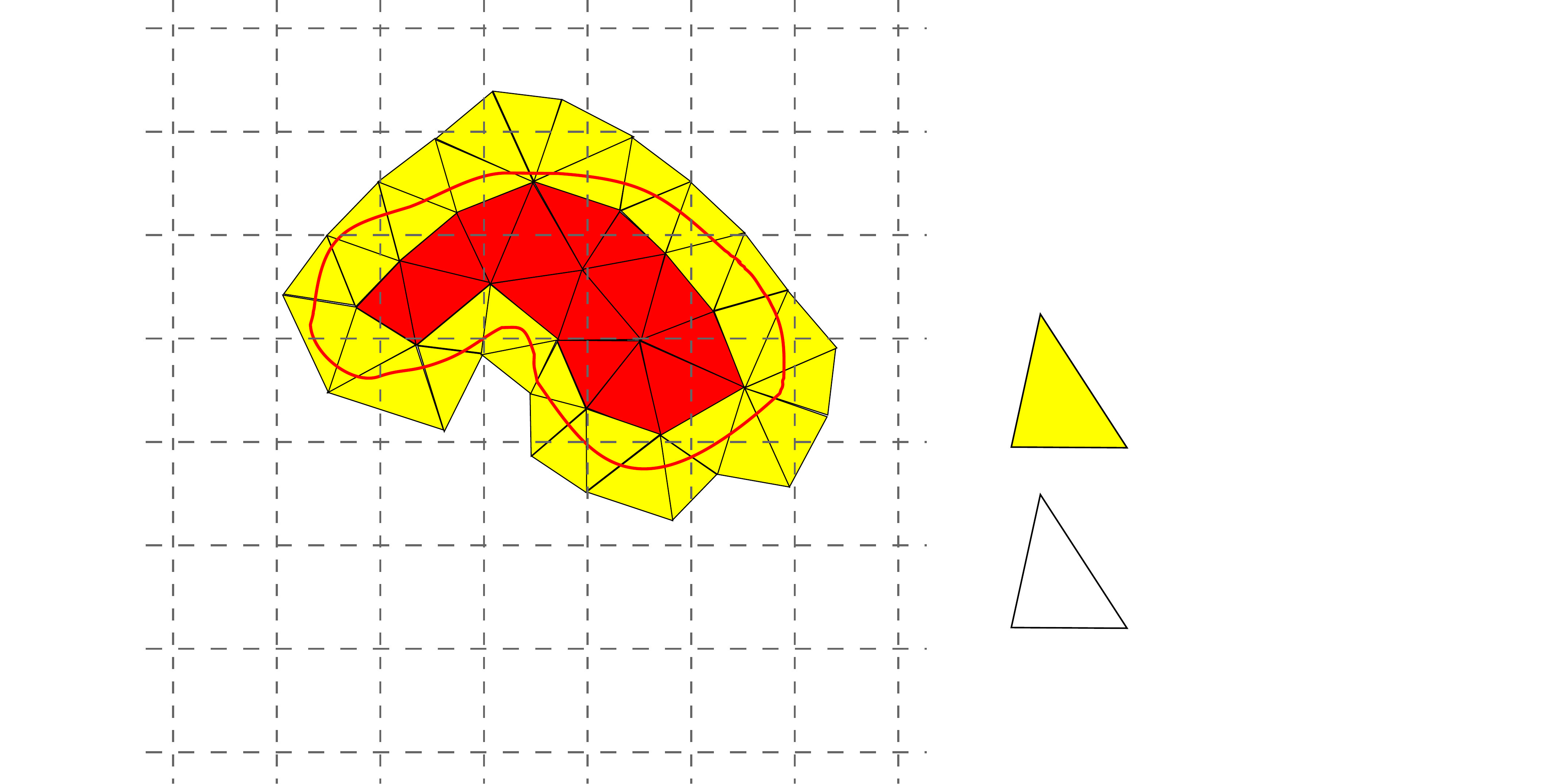
	  \caption{}
	  \label{fig:box2d_new}
      \end{subfigure}%
      	~ 
      \begin{subfigure}[b]{0.25\textwidth}
              \centering
	      \includegraphics[width=1\columnwidth]{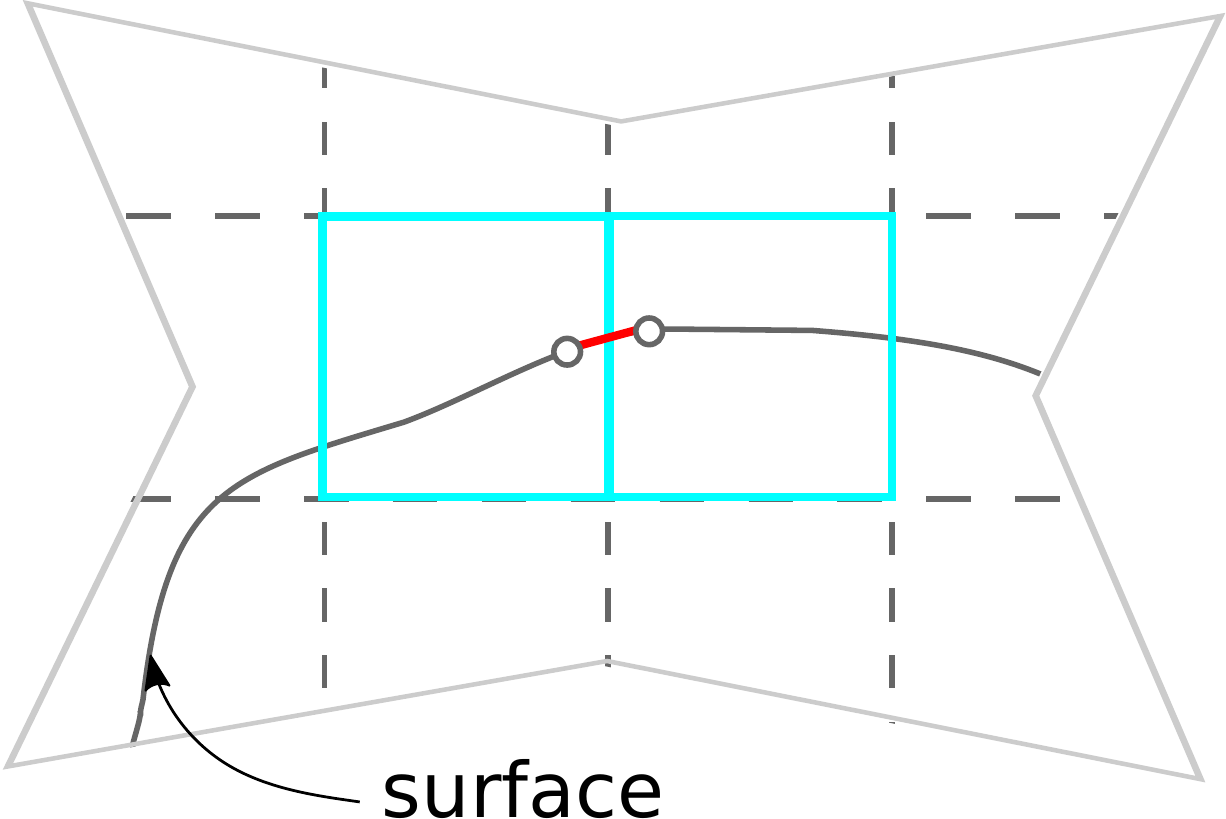}
              \caption{}
              \label{fig:cube_incident_to_triangle_on_surface}
      \end{subfigure}%
	~ 
      \begin{subfigure}[b]{0.15\textwidth}
              \centering
	      \includegraphics[width=1\columnwidth]{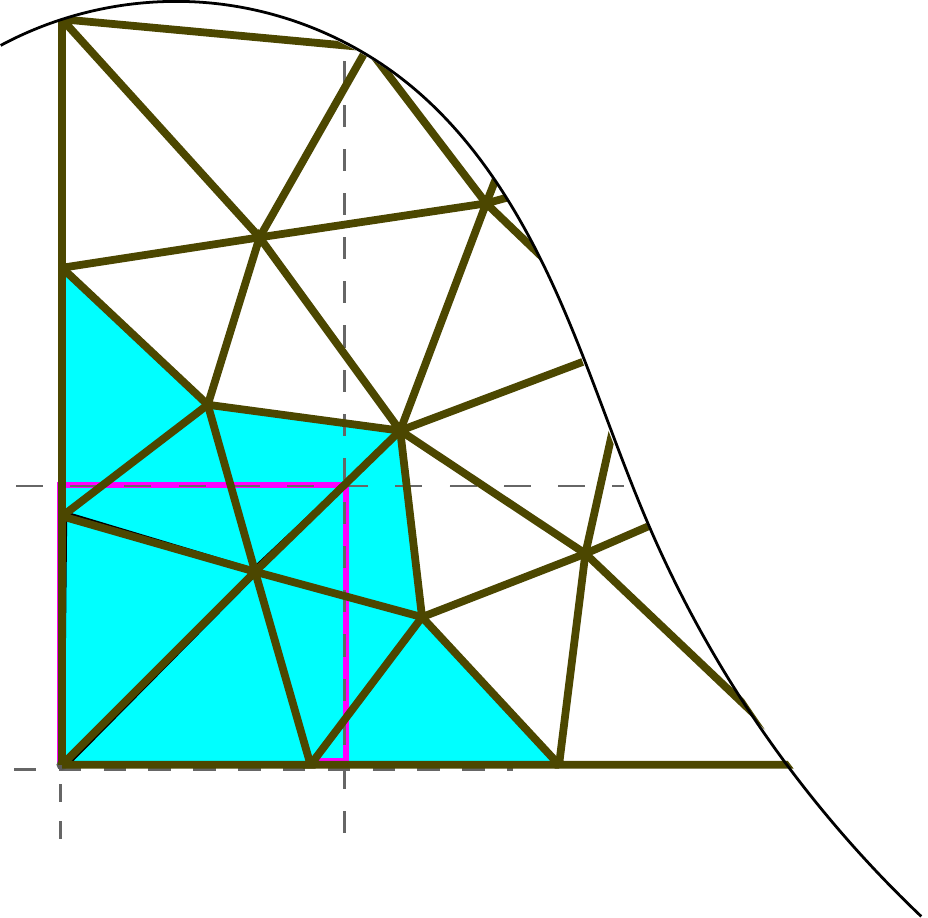}
              \caption{}
              \label{fig:elements_incident_to_cube}
      \end{subfigure}%
\caption{Subdivision of the bounding box (\subref{fig:box2d_new}), subcubes incident to each triangle of the interface are computed (\subref{fig:cube_incident_to_triangle_on_surface}), elements incident to a subcube are marked (\subref{fig:elements_incident_to_cube}).}\label{fig:boundingBox_subdivision}
\end{figure}

Since the interface surface is arbitrary and can be concave, only using level-set function to distinguish between elements that are fully contained in the volume bounded by the surface, elements that are completely outside that volume and elements are cut by the surface, may not be efficient due to the change of orientation of the outward normal of the surface. Moreover, for elements which are far from the surface, using the distance function as level-set function, in combination with outward normal of the surface, to mark elements, raises the ineffectiveness of the algorithm. To remedy that issue, we mark different types of element by the following technique, see also \cref{fig:box2d_new}.
{\ttfamily \small
\begin{itemize}
 \item[Step 1] Mark the cut elements firstly, labelled by $1$, by checking the potential intersections, using level-set function as described in \cref{sec:discretisationGeometry}, between each triangle of the surface mesh with \emph{only} elements (tetrahedra) surrounding the triangle,
 \item[Step 2] Mark the outside elements by propagating the marking procedure starting from the elements located at the boundary $\partial \Omega$ of the domain until a cut element is reached. During this checking propagation, all elements are marked as outside, labelled by $0$,
 \item[Step 3] Mark the remaining elements as inside, labelled by $2$.
\end{itemize}
}

To identify which tetrahedra are located around each triangle of the surface, used to check for potential intersections in Step 1, and to classify the tetrahedra located at the boundary domain $\partial \Omega$ for marking procedure to propagate from, in Step 2, we use the following approach. We first compute the bounding box of the domain, then the bounding box is subdivided into subcubes, as shown in \cref{fig:box2d_new}. All subcubes which are incident to each triangle of the surface, see \cref{fig:cube_incident_to_triangle_on_surface}, are then computed. All tetrahedra incident to each cube are also figured out, see \cref{fig:elements_incident_to_cube}. From these two data structures, one can easily access the tetrahedra around each triangle on the surface to check for intersections, and thus all cut elements can be marked. On the other hand, all tetrahedra located at the boundary $\partial \Omega$ can also be easily accessed from the boundary subcubes, to propagate marking procedure for outside elements. 

\begin{figure}[!htbp]
 \centering
\includegraphics[width=1\columnwidth]{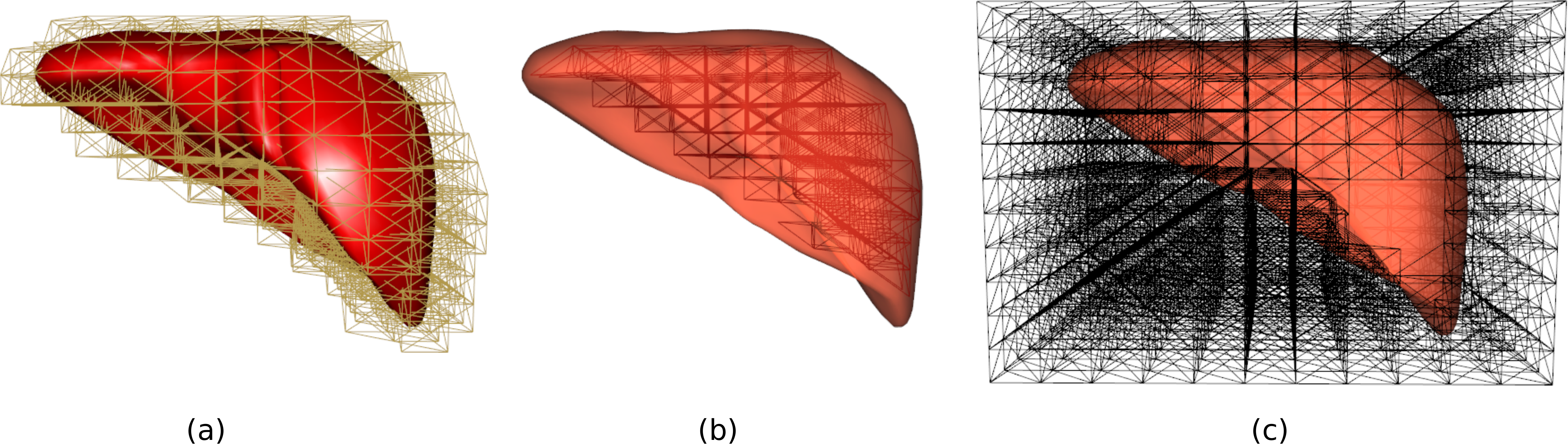}
\caption{Three type of tetrahedra are marked when a liver surface is immersed in a computational tetrahedron mesh: cut element (a), inside elements (b) and outside elements (c).}
\label{fig:liver_cut_inside_outside_elements}
\end{figure}

\cref{fig:liver_cut_inside_outside_elements} shows three type of tetrahedra marked when a liver surface is immersed in a tetrahedron mesh.

\begin{figure}[!htbp]
 \centering
       \begin{subfigure}[b]{0.3\textwidth}
	  \centering
	  \includegraphics[width=1\columnwidth]{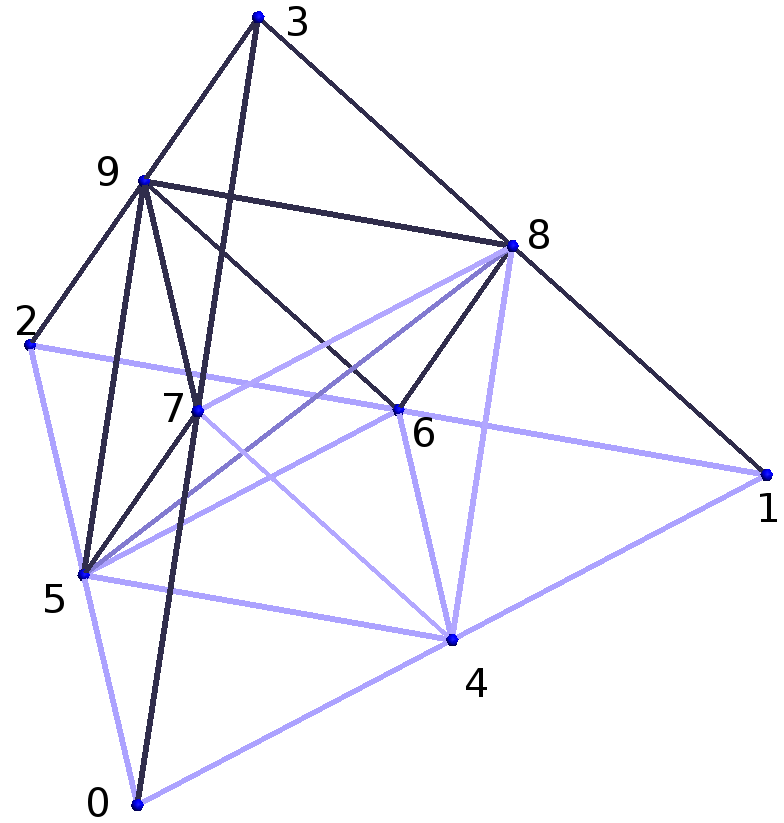}
	  \caption{}
	  \label{fig:template_8_tetrahedra_gmsh_cropped}
      \end{subfigure}%
      	~ 
      \begin{subfigure}[b]{0.65\textwidth}
              \centering
	      \includegraphics[width=1\columnwidth]{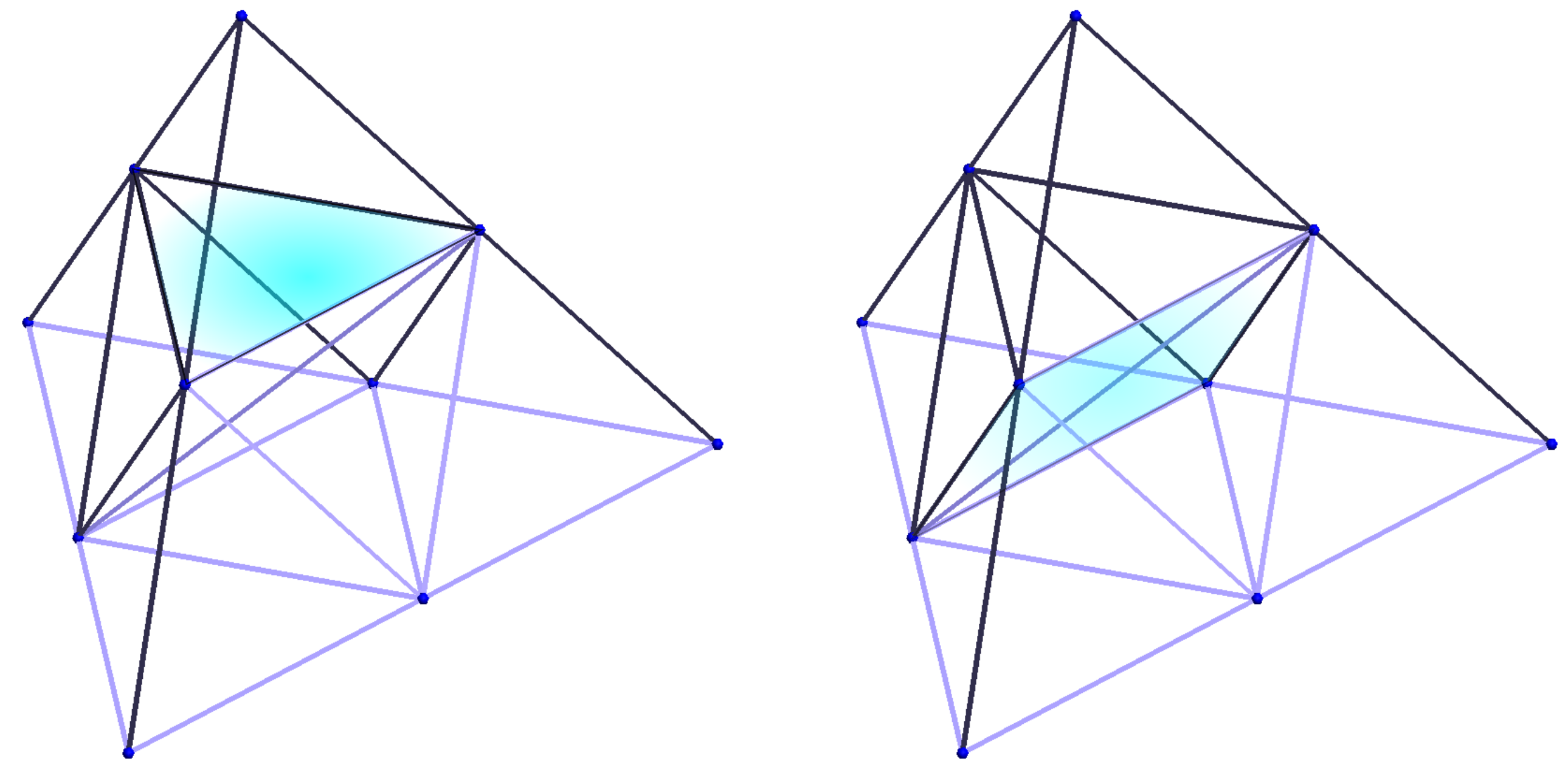}
              \caption{}
              \label{fig:template_cutting_cases}
      \end{subfigure}%
\caption{A set of eight template tetrahedra (\subref{fig:template_8_tetrahedra_gmsh_cropped}). Depending on whether the intersection between a surface and a cut tetrahedron is a triangular section or a quadrilateral one, the template is rotated and then mapped in the cut tetrahedron geometry (\subref{fig:template_cutting_cases}).}\label{fig:template_tetra_cutting_cases}
\end{figure}

Once all cut elements are identified, a set of subtetrahedra with conforming nodes regarding the interface surface is embedded in each cut tetrahedron to facilitate the integration. It is observed that there are only two kinds of intersection between a surface and a tetrahedron: with a triangle intersection or a quadrilateral one, as show in \cref{fig:tetrahedron_plane_intersection}. Therefore, it is sufficient to employ a set of tetrahedra, called the template set, as shown in \cref{fig:template_8_tetrahedra_gmsh_cropped}, to embed for each cut tetrahedron. Depending on each real case where the tetrahedron is cut by a surface with a triangular intersection or a quadrilateral one, the template set is rotated and then mapped into the cut tetrahedron geometry using the mapped mesh method \cite{Grosland2009,Bui2016_TBME}. It is noted that the template nodes 4, 5, 6, 7, 8, 9, see \cref{fig:template_8_tetrahedra_gmsh_cropped}, are located at the middle of their corresponding edges. If any edge is intersected by the interface surface, the corresponding midpoint is moved to match the real intersection between the cut tetrahedron and the interface, determined previously, as described in \cref{sec:discretisationGeometry}.


\begin{figure}[!htbp]
 \centering
       \begin{subfigure}[b]{0.47\textwidth}
	  \centering
	  \includegraphics[width=1\columnwidth]{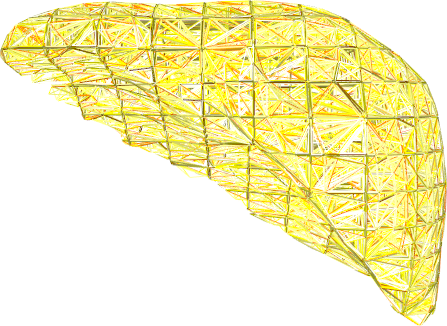}
	  \caption{}
	  \label{fig:liver_embedded_subtetrahedra_inside}
      \end{subfigure}%
      	~ 
      \begin{subfigure}[b]{0.53\textwidth}
              \centering
	      \includegraphics[width=1\columnwidth]{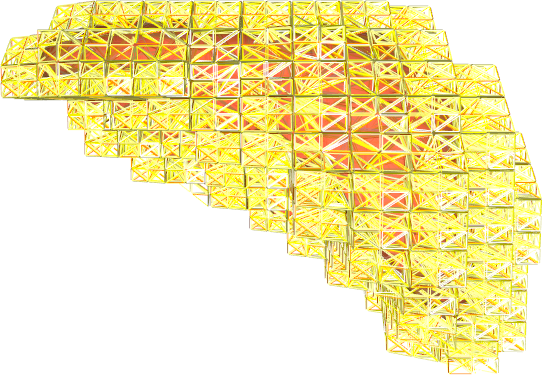}
              \caption{}
              \label{fig:liver_embedded_subtetrahedra_outside}
      \end{subfigure}%
\caption{Once a template tetrahedron set is embedded for each cut tetrahedron, subtetrahedra are marked as inside (\subref{fig:liver_embedded_subtetrahedra_inside}) or outside (\subref{fig:liver_embedded_subtetrahedra_outside}) with respect to the interface surface, using the level-set method.}\label{fig:liver_embedded_subtetrahedra_inside_outside}
\end{figure}

Once a template set is embedded for each cut tetrahedron, the last step consists in computing the relative location (inside or outside) of the subtetrahedra of the template with respect to the interface surface. The same level-set method is used. \cref{fig:liver_embedded_subtetrahedra_inside_outside} shows the subtetrahedra which are marked inside and outside with respect to the liver surface.

\begin{figure}[!htbp]
 \centering
\includegraphics[width=0.4\columnwidth]{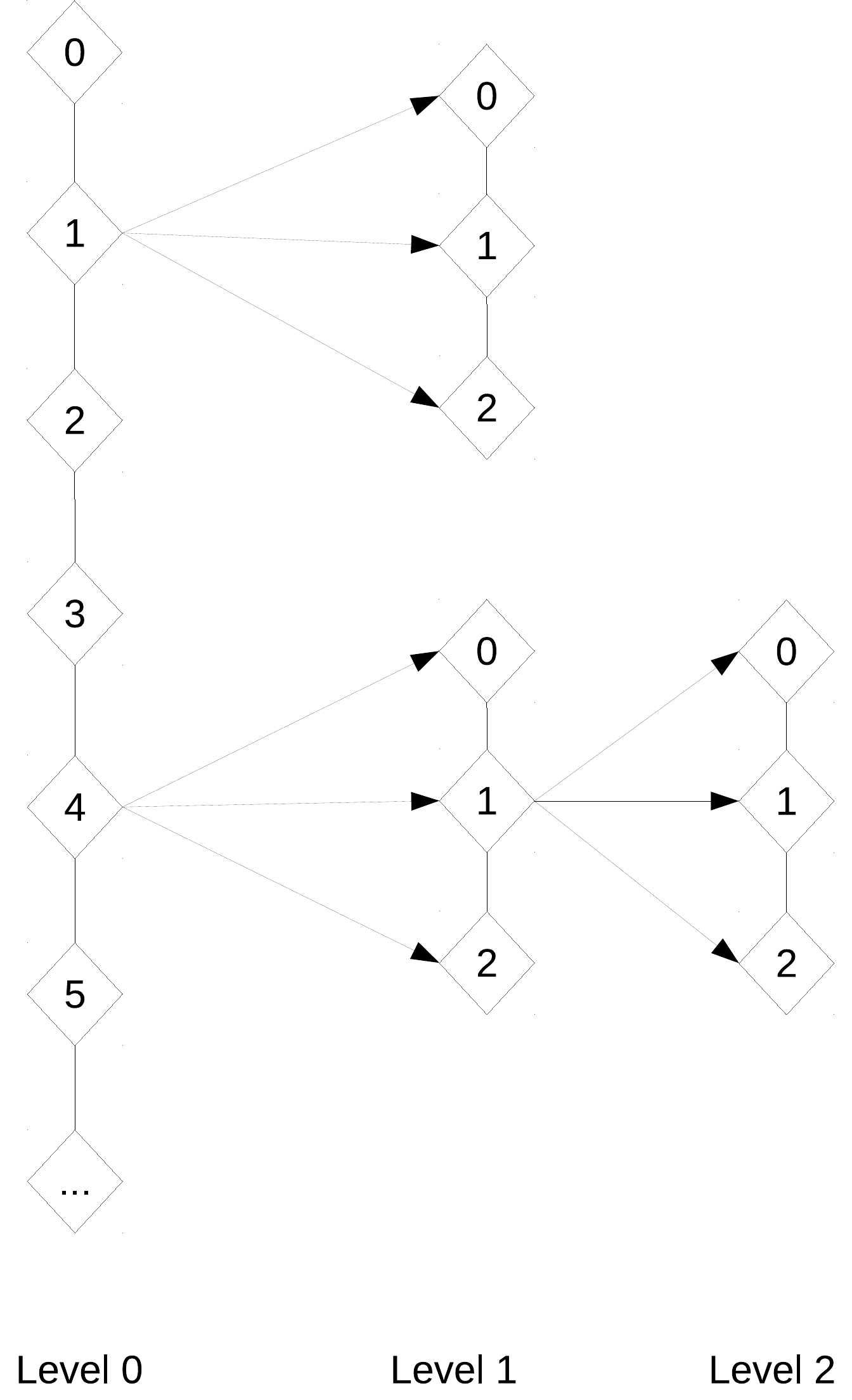}
\caption{Schematic representation of a tree data structure obtained when cut elements are embedded with a subtetrahedron set or when an invalid cut element is recursively refined to get all valid cut elements before embedding with a subtetrahedron set for integration reason. The elements of the background mesh are indexed at level 0, while embedded elements are indexed at subsequent levels.}
\label{fig:tree_data_structure}
\end{figure}

As described in \cref{sec:refinement_invalid_elements}, when an invalid cut element arises, it is recursively embedded with a tetrahedron set from a predefined template. It results in a tree data structure as schematically shown in \cref{fig:tree_data_structure}. To efficiently handle this type of data structure in implementation, we use a STL-like C++ tree class \url{http://tree.phi-sci.com/}. Using this container, it allows to recursively add embedded elements as children of the cut element (regarded as parent) very easily. The container also provides different kinds of iterators to access desired elements efficiently. 

\subsubsection{Numerical integration.}
\begin{figure}[!htbp]
  \centering
    \def\svgwidth{\textwidth}
    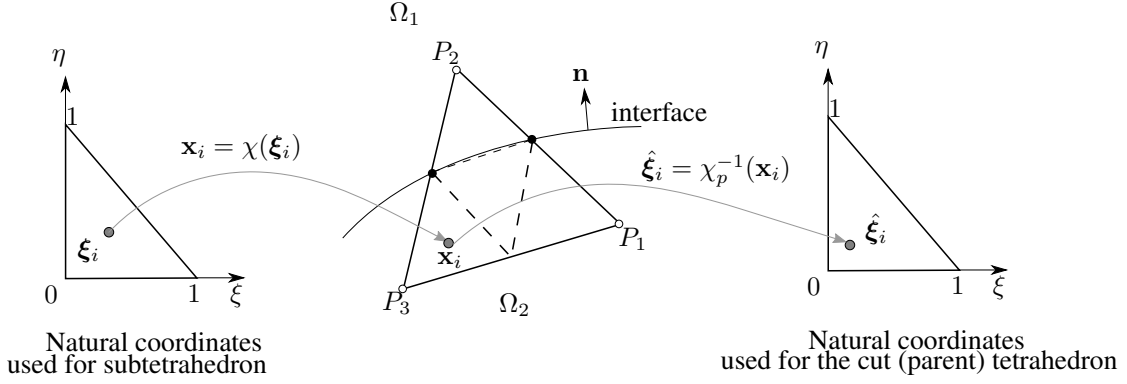
  \caption{Integration on cut elements. Note that the degrees of freedom are only defined on the element nodes $P_1$, $P_2$ and $P_3$. $\chi$ and $\chi_p$ are the coordinate mapping between the reference element with the physical sub-element and the cut (parent) element, respectively.}
  \label{fig:integration_on_cut_element}
\end{figure}

For those elements that are fully contained in the domain $\Omega_1$ or $\Omega_2$, see \cref{fig:domain_definition}, integration of element stiffness and mass matrices are performed normally as in classical FEM. Only for cut elements, the integration is split into two parts which are related to the inside and outside subtetrahedra regarding the interface. A two dimensional schematic representation of the integration using natural coordinates on reference element is shown in \cref{fig:integration_on_cut_element}.

The stiffness matrix of the cut element reads
\begin{equation}
\b{K}_e = \b{K}_e^{\Omega_1} + \b{K}_e^{\Omega_2},
\end{equation}
where $\b{K}_e^{\Omega_1}$ and $\b{K}_e^{\Omega_2}$ denote the stiffness contributions of the part belonging to $\Omega_1$ and $\Omega_2$, respectively, to the element $e$. The stiffness matrix on each part is computed by summing the contributions from their sub-elements. For example, the stiffness matrix $\b{K}_e^{\Omega_2}$ reads

\begin{equation}
\b{K}_e^{\Omega_2} = \sum_{\Omega_2^k} \int_{\Omega_2^k} \b{B}_s^T \b{E}_2 \b{B}_s \, \mathrm{d}\Omega_2^k =\sum_{\Omega_2^k} \sum_{i=1}^{N_p} \b{B}_s(\bm{\xi}_i )^T \b{E}_2 \b{B}_s(\bm{\xi}_i ) \omega_i \text{det}(\b{J}(\bm{\xi}_i)).
\label{eq:KeOmega2}
\end{equation}
in which $\b{B}_s$ is the strain displacement matrix of the sub-element $s$,  $\b{E}_2$ is the material stiffness tensor of the domain $\Omega_2$,  $\bm{\xi}_i$ and $\omega_i$ are the quadrature coordinates and the corresponding weight parameters, $N_p$ is the number of quadrature points used, $\b{J}$ is the Jacobian matrix of the coordinate transformation. Since the stiffness matrix should be expressed on the cut (parent) element where the degrees of freedom are defined, we must compute the strain displacement matrix $\b{B}_p$ of the parent element at the quadrature point $\hat{\bm{\xi}}_i$ corresponding to the physical coordinates $\b{x}_i$ of the sub-element, see \cref{fig:integration_on_cut_element}. So, \cref{eq:KeOmega2} reads

\begin{equation}
 \b{K}_e^{\Omega_2} = \sum_{\Omega_2^k} \sum_{i=1}^{N_p} \b{B}_p(\hat{\bm{\xi}}_i )^T \b{E}_2 \b{B}_p(\hat{\bm{\xi}}_i ) \omega_i \text{det}(\b{J}(\hat{\bm{\xi}}_i)).
\end{equation}
with $\b{B}_p$ the strain displacement matrix defined on the parent element $P_1P_2P_3$.

In this paper, we use linear tetrahedra, the strain displacement matrix is constant across the element volume, and $\omega_i= 1/6$, $\text{det}(\b{J}) = 6 V_k$ with $V_k$ the sub-element volume, see \eg{} \cite{Dhatt2012} for more details, one gets 
\begin{equation}
 \b{K}_e^{\Omega_2} = \sum_{\Omega_2^k} \b{B}_p^T \b{E}_2 \b{B}_p V_k.
\end{equation}
The computation for $\b{K}_e^{\Omega_1}$ can be done using the same concept. Also, integration over cut elements for the mass matrices is performed by the same procedure.

\subsection{Corotational formulation for cut FEM}
\label{sec:corotationalFormulation}
%

In many surgical simulations, tissues undergo large displacements and rotations, see \eg{} \cite{Plantefeve2016,Bui_2017_error_brain}. Using linear elasticity for modelling of soft tissues results in artifacts for large rotational deformation \cite{muller2002stable}. To overcome this issue, the stiffness matrix is computed based on the corotational formulation \cite{Felippa2005}, in which the rigid motion can be extracted from the total finite element displacements. The element nodal internal force becomes
\begin{equation}
 \mathbf{f}_e = \mathbf{R}_e \mathbf{K}_e ( \mathbf{R}^T_e \mathbf{x}_e - \mathbf{x_0}_e),
\end{equation}
where $\mathbf{R}_e$ stands for the element rotation matrix of the element local frame with respect to its initial orientation, being updated at each time step.

The element rotation matrix is computed from the deformation gradient, using polar decomposition. As for deformation gradient $\b{F}$, it is computed as
\begin{equation}
 \b{F} = \b{P}_n \cdot \b{N'},
\end{equation}
with $\b{P}_n$ the element nodal coordinates, $\b{N'}$ the derivative of shape functions. For tetrahedron element, $\b{P}_n$ and $\b{N'}$ are $3 \times 4$ and $4 \times 3$ matrices, respectively. Employing polar decomposition, one gets the rotation matrix $\b{R}$ from
\begin{equation}
 \b{F} = \b{R} \cdot \b{U},
\end{equation}
with $\b{U}$ is the right stretch tensor that is responsible for tissue deformation.


\subsection{Boundary conditions on immersed surfaces}
\label{sec:boundary_condition_on_immersed_surfaces}





\begin{figure}[!htbp]
  \centering
    \def\svgwidth{0.8\textwidth}
    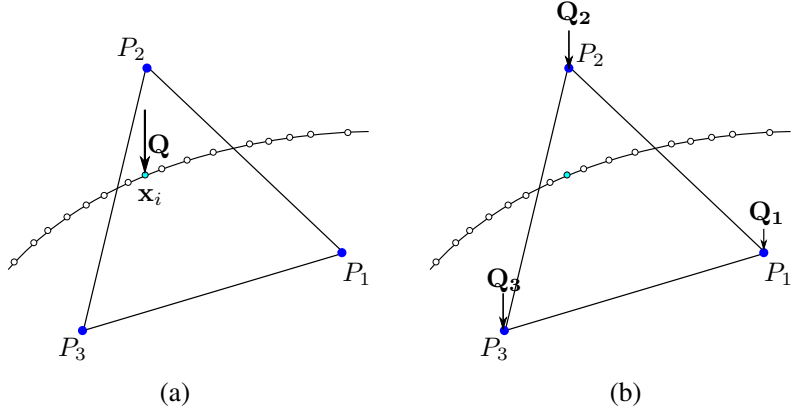
  \caption{A force $\b{Q}$ applied on the surface at point $\b{x}_i$ (a), is barycentrically mapped into the nodes of the element from the computational mesh containing $\b{x}_i$ (b).}
  \label{fig:neumann_boundary_condition}
\end{figure}

When using immersed surface, one needs to impose Neumann and Dirichlet boundary conditions on the immersed surface, which does not conform with the computational mesh. For the Neumann boundary condition, a force applied on the surface is barycentrically mapped into the nodes of the element from the computational mesh which contains the applied point of that force, see \cref{fig:neumann_boundary_condition}.

The concept of this approach is based on the master-slave scheme \cite{Rabczuk2010_immersedParticle} in which the displacement of a point on the surface can be seen to be mapped by the displacement of the computational mesh as
\begin{equation}
 \b{u}_S = \b{J} \b{u}_M,
 \label{eq:us=Jum}
\end{equation}
with $\b{u}_S$ the displacement of the surface (is considered as \emph{slave}), $\b{u}_M$ the displacement of the computational mesh (is regarded as \emph{master}), and $\b{J}$ the barycentric coordinates of the considered point on the surface with respect to the computational mesh. By applying the principle of virtual work
\begin{equation}
 \b{u}_M^T \b{Q}_M = \b{u}_S^T \b{Q}_S,
 \label{eq:virtualWork}
\end{equation}
where $\b{Q}_S$, $\b{Q}_M$ are the forces applied on the surface and the equivalent forces applied at the computational mesh. By substituting \cref{eq:us=Jum} into \cref{eq:virtualWork}, one obtains
\begin{equation}
 \b{Q}_M = \b{J}^T \b{Q}_S.
\end{equation}


Dirichlet boundary conditions at some nodes $i$ on the surface, can be imposed as $\b{u}_S^i = \bar{\b{u}}$ with $\bar{\b{u}}$ the prescribed displacement. Taking into account \cref{eq:us=Jum}, the Dirichlet boundary condition can be expressed as
\begin{equation}
 \b{J} \b{u}_M = \bar{\b{u}}.
\end{equation}
Using Lagrange multipliers, one can easily impose the prescribed displacement on the computational mesh by solving the system equation set
\begin{equation}
\label{eq:Dirichlet}
 \begin{pmatrix}
  \mathbf{A} & \mathbf{J}^T \\[0.3em]
  \mathbf{J} & \mathbf{0}
 \end{pmatrix}
 \begin{Bmatrix}
                \mathbf{u}_M \\[0.3em]
                \bm{\lambda}
\end{Bmatrix}
               = \begin{Bmatrix}
                  \mathbf{b} \\[0.3em]
                  \mathbf{0}
                 \end{Bmatrix},
\end{equation}
where $\bm{\lambda}$ stands for Lagrange multipliers used for Dirichlet boundary conditions.

\subsection{Solving system equations with constraints}
\label{sec:solveSystem}



The interaction between the needle (denoted by subscript $1$) and the tissue (denoted by subscript $2$) can be expressed by the following equation set
\begin{equation}
\label{eq:needle_tissue_interaction}
 \begin{pmatrix}
  \b{A}_1 & \b{0} & \b{H}_1^T \\[0.3em]
  \b{0} & \b{A}_2 & \b{H}_2^T \\[0.3em]
  \b{H}_1 & \b{H}_2 & \b{0}
 \end{pmatrix} \begin{Bmatrix}
                d\b{v}_1 \\[0.3em] d\b{v}_2 \\[0.3em] \bm{\lambda}_i
               \end{Bmatrix}
               = \begin{Bmatrix}
                  \b{b_1} \\[0.3em] \b{b_2} \\[0.3em] \b{0}
                 \end{Bmatrix},
\end{equation}
where $\bm{\lambda}_i$ is the Lagrange multiplier representing the \emph{interaction} between the needle and the tissue.
We can see that, \cref{eq:needle_tissue_interaction} describing the interaction between the needle and the tissue, and \cref{eq:Dirichlet} expressing the constraints used for Dirichlet condition on implicit boundaries, have the same general form
\begin{equation}
\label{eq:generalConstraintEqs}
 \begin{pmatrix}
  \b{A} & \b{J}^T \\[0.3em]
  \b{J} & \b{0}
 \end{pmatrix}
 \begin{Bmatrix}
                \b{x} \\[0.3em]
                \bm{\lambda}
\end{Bmatrix}
               = \begin{Bmatrix}
                  \b{b} \\[0.3em]
                  \b{0}
                 \end{Bmatrix}.
\end{equation}
\cref{eq:generalConstraintEqs} can be reformulated as
\begin{subequations}
 \begin{align}
  \b{x} & = \underbrace{\b{A}^{-1} \b{b} }_{\b{x}_{free}} - \b{A}^{-1} \b{J}^T \bm{\lambda}, \label{eq:x} \\
  \b{J} \b{A}^{-1} \b{J}^T \bm{\lambda} & =  \b{J} \underbrace{\b{A}^{-1} \b{b} }_{\b{x}_{free}}, \label{eq:lambda}
 \end{align}
\end{subequations}
in which, $\b{x}_{free}$ can be seen as the solution of the unconstrained system $\b{A} \b{x} = \b{b}$. Therefore, \cref{eq:generalConstraintEqs} can be solved in three steps as
{\ttfamily \small
\begin{enumerate}
\setlength{\itemindent}{1em}
 \item[Step 1] Factorise the matrix $\b{A}$ to have its inverse $\b{A}^{-1}$, and solve for $\b{x}_{free}$,
 \item[Step 2] Solve Lagrange multipliers $\bm{\lambda}$ from \cref{eq:lambda},
 \item[Step 3] Once $\bm{\lambda}$ is available, $\b{x}$ can be obtained from \cref{eq:x} by using $\b{x}_{free}$.
\end{enumerate}
}

\section{Results}
\label{sec:Results}

\subsection{Convergence study}
\label{sec:convergence_study}

\begin{figure}[!htbp]
 \centering
\includegraphics[width=0.6\columnwidth]{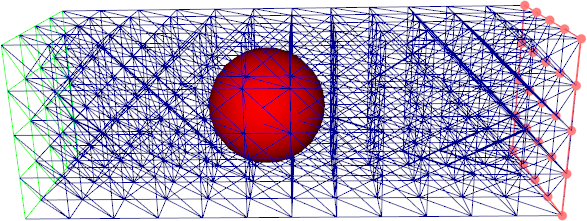}
\caption{A beam is fixed at right end (marked by the red points), and is subjected to, at left end (marked by the green triangles), a uniform horizontal pressure (the tensile case) or a uniform vertical pressure (for bending case). A sphere surface is immersed inside the beam geometry.}
\label{fig:beam_study}
\end{figure}

In order to verify the implementation of the cut FEM, we study the convergence of a tensile test and a bending one under mesh refinement. A beam, with a sphere surface being embedded inside, which is subjected to, at one end, a uniform horizontal pressure (tensile) or a uniform vertical pressure (bending), while other end being clamped, as shown in \cref{fig:beam_study}, is studied. For the convergence study, the linear elastic constitutive law is used. The mechanical properties for the material outside the sphere are $E_1$, $\nu_1$ being the Young's modulus and the Poisson's ratio, respectively. Those for the material inside the sphere are $E_2$ and $\nu_2$. The dimension of the beam is $6 \times 2 \times 2$. In order to compare rate convergences of the tests with the theoretical ones, we use the same mechanical properties for the material inside and outside the sphere surface. We set thus $E_1=E_2 = 1000$ and $\nu_1 = \nu_2 = 0.1$.

The convergence is studied by computing the solution of the tensile and bending tests employing the tetrahedral meshes consisting of $7 \times 3 \times 3$, $13 \times 5 \times 5$, $25 \times 9 \times 9$, $49 \times 17 \times 17$ nodes. We propose to use the solution from the classical FEM when employing a very fine mesh ($97 \times 33 \times 33$ nodes) as the reference solution. We then study the convergence of the error between the cut FEM solution and the reference FEM solution. This error is measured by using both the $L_2$ norm and the energy one. The $L_2$ norm of the displacement error reads

\begin{equation}
 \lVert \eta \rVert_{L_2} = \sqrt{ \frac{\int_\Omega (u_h - u_r)^2 \, \mathrm{d} \Omega }{ \int_\Omega u_r^2 \,  \mathrm{d} \Omega} },
\end{equation}
with $u_h$ the displacement solution of the cut FEM, and $u_r$ the reference solution of the classical FEM. About the energy norm, it is defined as
\begin{equation}
  \lVert \eta \rVert_{Energy} = \sqrt{ \frac{\int_\Omega (\sigma_h - \sigma_r) \cdot (\epsilon_h - \epsilon_r) \, \mathrm{d} \Omega }{ \int_\Omega \sigma_r \cdot \epsilon_r \, \mathrm{d} \Omega} },
\end{equation}
in which $\sigma_h$, $\epsilon_h$ denote the stress and strain of the cut FEM, respectively, and $\sigma_r$, $\epsilon_r$ stand for the reference values.

\begin{figure}[!htbp]
 \centering
       \begin{subfigure}[b]{0.5\textwidth}
	  \centering
	  \includegraphics[width=1\columnwidth]{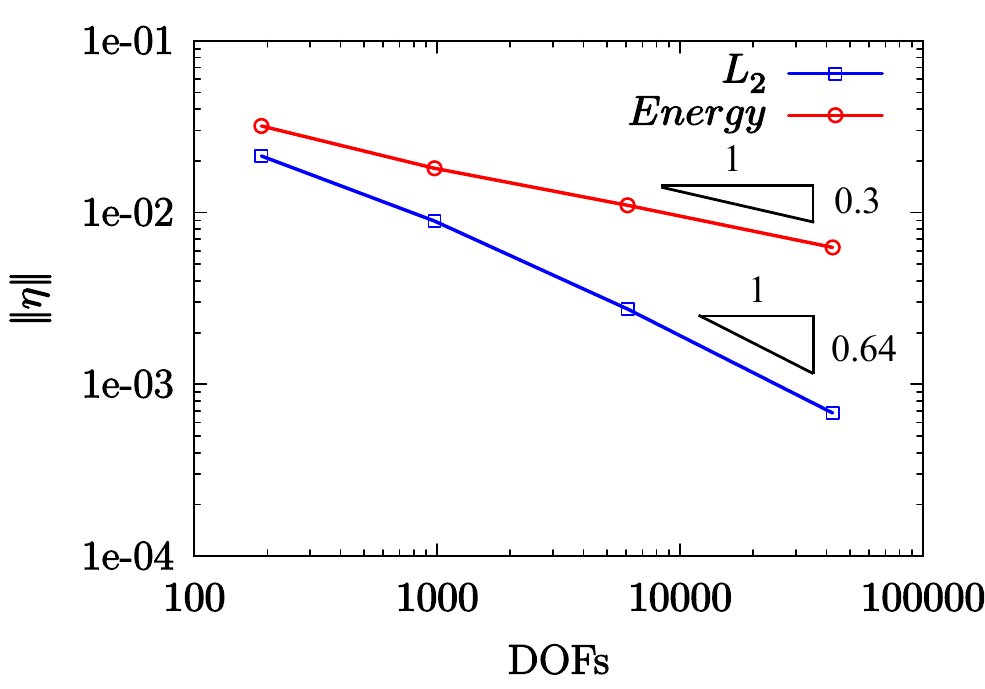}
	  \caption{}
	  \label{fig:traction_test_error_plot_slope}
      \end{subfigure}%
      	~ 
      \begin{subfigure}[b]{0.5\textwidth}
              \centering
	      \includegraphics[width=1\columnwidth]{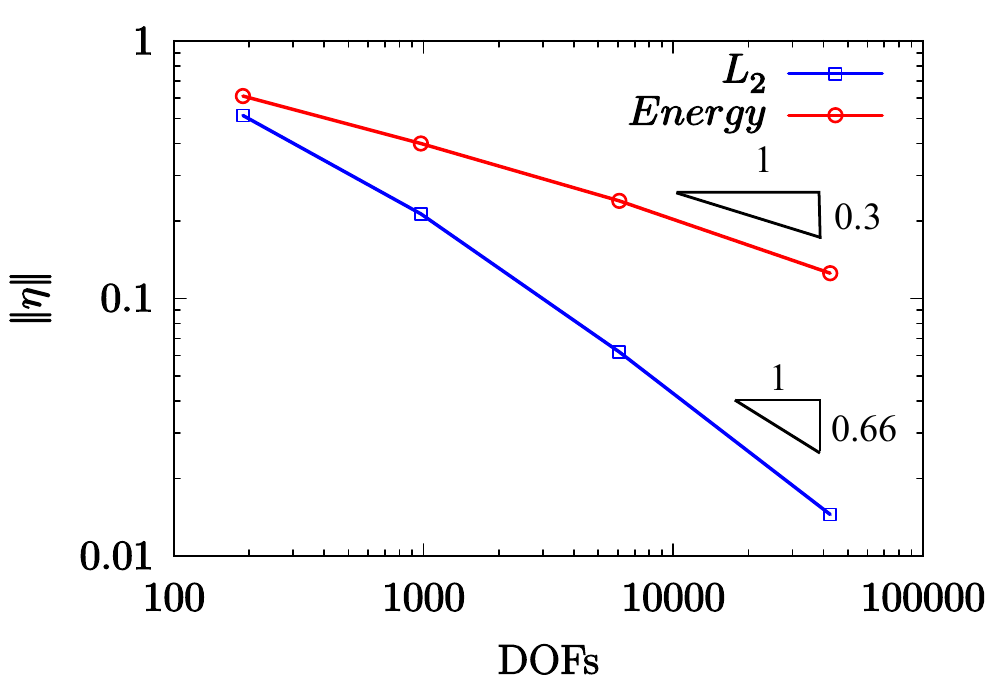}
              \caption{}
              \label{fig:bending_test_error_plot_slope}
      \end{subfigure}%
\caption{Convergence rates under mesh refinement for the tensile test (\subref{fig:traction_test_error_plot_slope}) and for the bending test (\subref{fig:bending_test_error_plot_slope}).}\label{fig:convergenceRate}
\end{figure}

\cref{fig:convergenceRate} shows the $L_2$ and energy norms versus the number of DOFs of the tensile and bending tests, also with the convergence rate. It is observed that the rates of convergence of the $L_2$ and energy norms for both tensile and bending tests agree well with the theoretical rates. Indeed, for 3D problems using linear elements, $L_2$ norm of displacement error is of the order $O(N^{-2/3})$, while energy norm converges with an order of $O(N^{-1/3})$, with $N$ the total number of the degrees of freedom.

\subsection{A comparison with FEM}
The aim is to compare simulation results obtained from the cut FEM using non-conforming meshes with those obtained from the classical FEM using conforming meshes. The displacement is considered here for comparison. We carry out the study on two different geometries: the simple beam geometry with a immersed sphere as shown in \cref{fig:beam_study}, and the more complex liver geometry shown in \cref{fig:liver_in_tetrahedron_mesh_cropped}.

\begin{figure}[!htbp]
 \centering
\includegraphics[width=0.8\columnwidth]{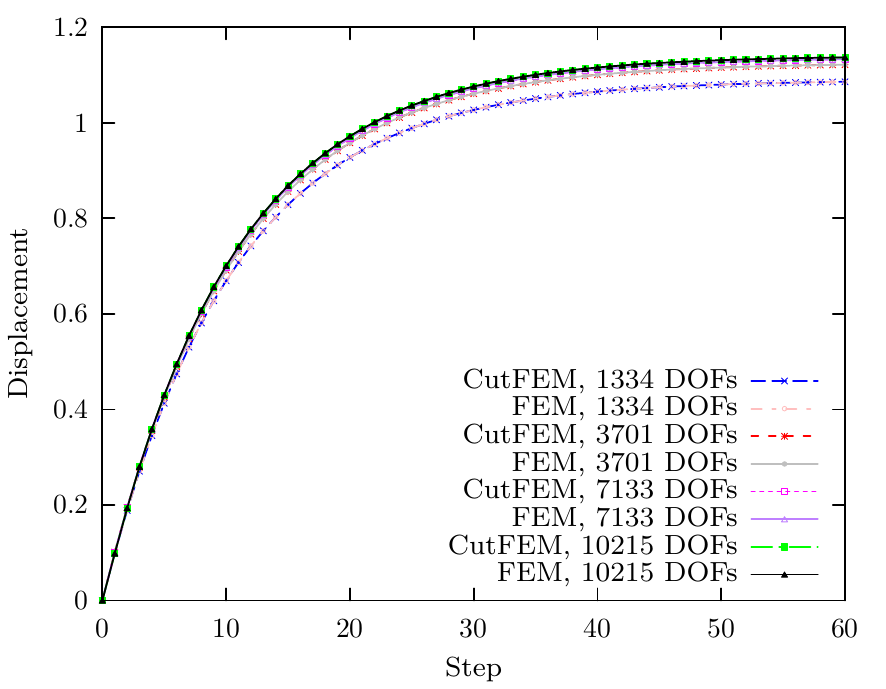}
\caption{Displacement measured at the centre of the left end of the beam during simulations by the cut FEM and the classical FEM.}
\label{fig:beam_displacement}
\end{figure}

For the beam geometry, the same dimensions, and the mechanical properties, as in \cref{sec:convergence_study} are used. The displacement measured at the centre of the left end of the beam is employed to compare between the cut FEM (where the sphere is modelled implicitly) and the classical FEM. The beam is subjected to a uniformly distributed pressure $q=-5$ in the vertical direction at the left end, whereas being clamped at the right end. \cref{fig:beam_displacement} shows the displacement of the point during simulations using the cut FEM and the classical FEM, under mesh refinement. It is observed that, at the same number of degrees of freedom used, the result obtained from the cut FEM perfectly agrees with that of the FEM. Also, under mesh refinement, the displacement asymptotically convergences to the solution of the fine mesh. 

\begin{figure}[!htbp]
 \centering
\includegraphics[width=0.5\columnwidth]{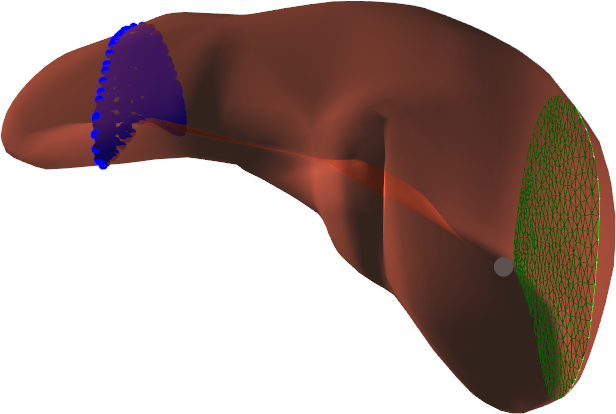}
\caption{Liver is implicitly clamped at the points shown in blue colour, and is implicitly subjected to a uniformly distributed vertical pressure acting on the mesh shown in green colour. The displacement is measured at the point, shown by the grey colour, during the simulations.}
\label{fig:liver_boundary_conditions}
\end{figure}

For the liver, due to its complex geometry, in order to apply the same boundary conditions acting on different conforming meshes (used for FEM), and on different non-conforming ones (used for the cut FEM), an homogeneous Dirichlet boundary condition is implicitly applied to the model through the points located on an imaginary cutting section, shown by the points in blue colour in \cref{fig:liver_boundary_conditions}, whereas an uniformly distributed pressure $q=-1$ is implicitly applied to the model in the vertical direction through the mesh shown by green colour in \cref{fig:liver_boundary_conditions}. The displacement is measured at the point inside the liver shown by the grey colour in \cref{fig:liver_boundary_conditions}. Young's modulus and Poisson's ratio, used for the simulations, are $1000$, and $0.4$ respectively.

\begin{figure}[!htbp]
 \centering
\includegraphics[width=0.8\columnwidth]{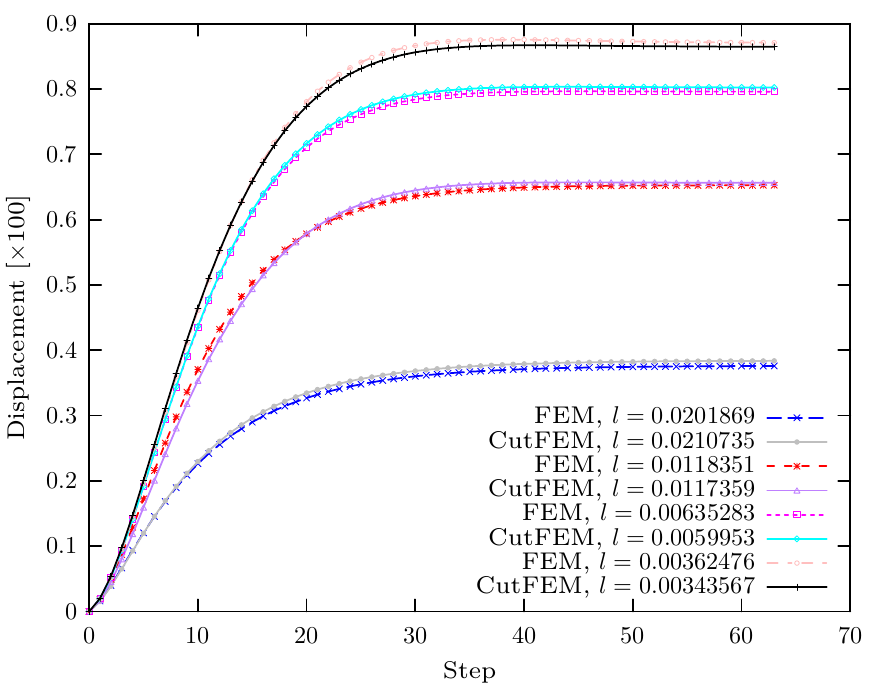}
\caption{Displacement measured at the point shown in \cref{fig:liver_boundary_conditions} during simulations by the cut FEM, and by FEM, with different mesh resolutions.}
\label{fig:liver_displacement}
\end{figure}

\cref{fig:liver_displacement} points out the displacement of the point shown in \cref{fig:liver_boundary_conditions}. It is observed that, at the same mean size of the elements used, the results obtained from the cut FEM agrees well with those of the FEM. It is noted that the number of degrees of freedom is not used as a the same input during the comparison between the cut FEM and FEM since it does not characterise the same mesh resolution between conforming (used for FEM) and non-conforming (used for cut FEM) meshes, with respect to the liver geometry. Instead, the mean element size $l$ is employed (see \cref{fig:liver_displacement}).

\subsection{Needle insertion simulations}




The aim is to employ the cut FEM approach and apply for needle insertion problems.

\subsubsection{Immersed interface.}

\begin{figure}[!htbp]
 \centering
\includegraphics[width=0.75\columnwidth]{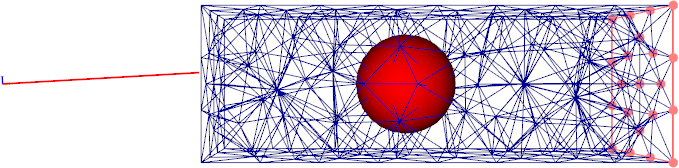}
\caption{Schematic illustration of a needle insertion simulation into a simple tissue geometry. The needle is initially inclining at an angle of $3.5$ degrees.}
\label{fig:needle_beam/geometry}
\end{figure}

The needle, which is initially inclining at an angle of $3.5$ degrees, is inserted into a phantom tissue with a simple geometry, as shown in \cref{fig:needle_beam/geometry}. We simulate a spherical inclusion (can be seen as a tumour) which is implicitly immersed in the tissue phantom. The dimension of the tissue phantom is $6 \times 2 \times 2$, while the radius of the inclusion is $0.7$. The length of the needle is $2.8$ and its cross section radius is $0.05$. Young's modulus and Poisson's ratio of the needle is set to $20\,000$ and $0.2$, respectively. These parameters for the phantom tissue are $E_1 = 1\, 000$ and $\nu=0.4$. Poisson's ratio of the inclusion is also set to $0.4$ whereas its Young's modulus $E_2$ is varied with respect to that of the phantom tissue by a factor of $1$, $2$, $4$ and $8$, in order to investigate the effect of the inclusion stiffness on the needle-tissue interaction force profile.

The penetration strength at the tissue surface is set to $1$, and the frictional coefficient between the tissue and the needle shaft is set to $0.5$.

\begin{figure}[!htbp]
 \centering
\includegraphics[width=0.75\columnwidth]{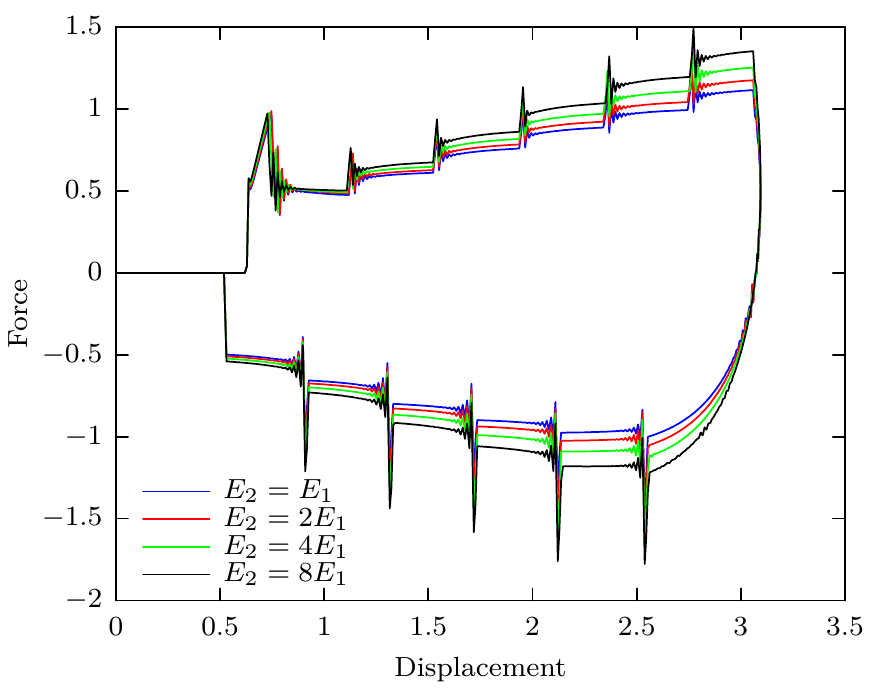}
\caption{Force displacement curves with varying Young's modulus ratio $E_2/E_1$ when the needle is inserted into the soft tissue model. When the needle tip displacement reaches about $3$, the needle is retracted, and thus generating negative needle-tissue interaction force; at the stage when the needle is completely retracted from the tissue, the interaction force gets zero.}
\label{fig:needle_beam/force_displacement}
\end{figure}

\cref{fig:needle_beam/force_displacement} shows the needle-tissue interaction force with respect to the displacement of the needle tip, with different ratios $E_2/E_1$. It is observed that when the needle tip reaches the tissue surface, the interaction force between the needle and the tissue occurs. This interaction force continuously increases and when it reaches the tissue surface penetration strength, the needle penetrates into the tissue. It also reveals that the closer to the inclusion the needle tip is, the more different the interaction force profiles when varying the ratio $E_2/E_1$ are obtained. This is indeed logical due to the higher stiffness of the inclusion compared to that of the tissue. When the displacement of the needle tip reaches $3$, the needle is continuously retracted until completely outside the tissue. During retraction process, the interaction force changes its sign and is negative, as observed in \cref{fig:needle_beam/force_displacement}. The same conclusion about the effect of inclusion-tissue stiffness ratio on the interaction force profile, during retraction phase can be drawn as during the insertion stage.

\subsubsection{Fictitious boundary.}

\begin{figure}[!htbp]
 \centering
\includegraphics[width=0.8\columnwidth]{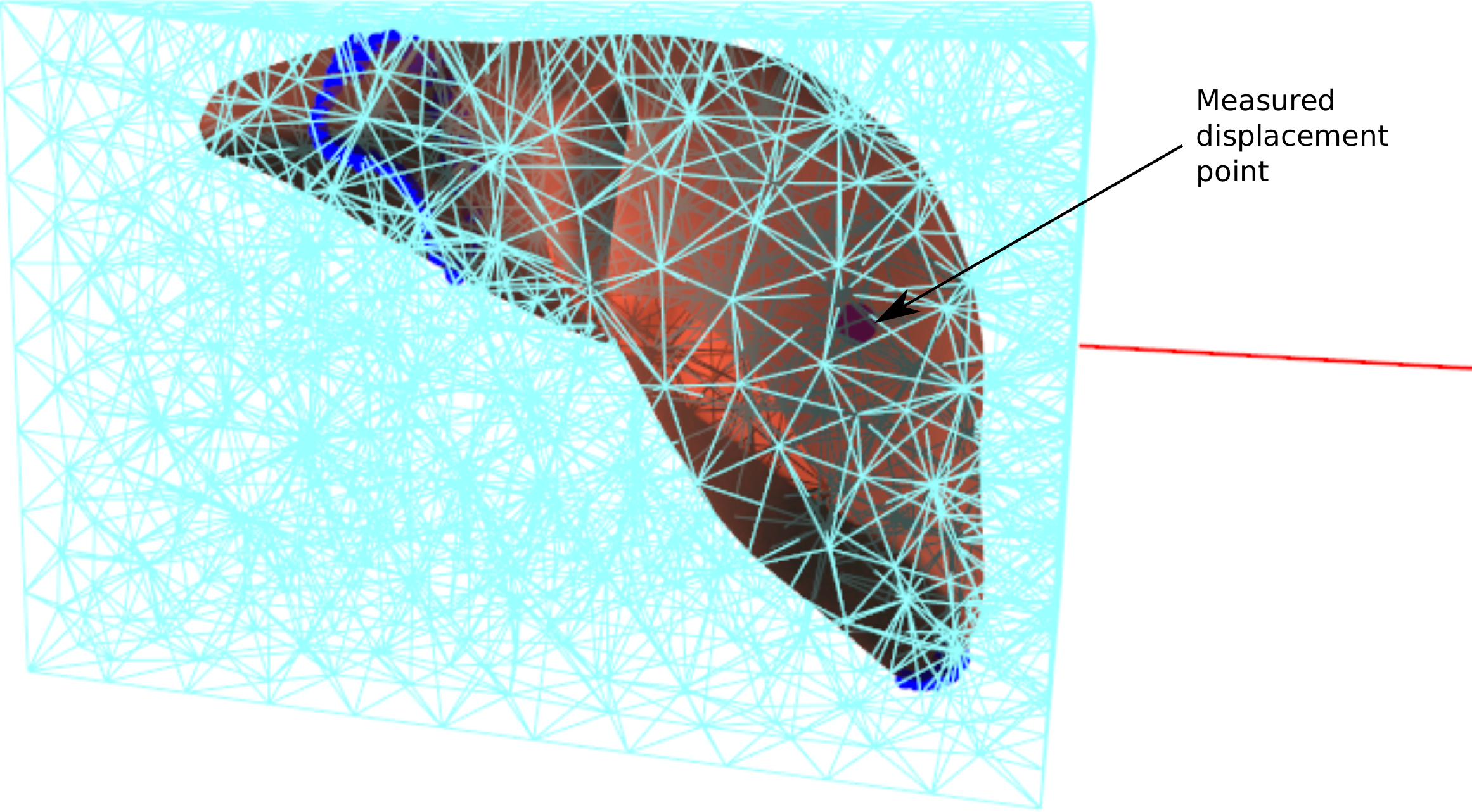}
\caption{Background mesh used for simulation of the liver behaviour during needle insertion and retraction. Some constraints are defined implicitly on the liver surface, using Lagrange multipliers, shown by the blue points. During simulation, displacement is measured at the point, located near to the needle shaft, shown in the figure.}
\label{fig:liver_needle_backgroundMesh}
\end{figure}

The focus is now on providing a comparison between a needle insertion simulation into a liver using fictitious (implicit) boundaries, and with those using explicit (conforming) boundaries. Young's modulus and Poisson's ratio are set for the liver are $1\, 200$ kPa and $0.4$, respectively. The same parameters for the needle as above are used. The background mesh used for simulation with CutFEM, and the boundary conditions applied to the liver surface, are shown in \cref{fig:liver_needle_backgroundMesh}. During the insertion and retraction of the needle, the displacement is measured at the point shown also in \cref{fig:liver_needle_backgroundMesh}.

\begin{figure}[!htbp]
 \centering
\includegraphics[width=0.75\columnwidth]{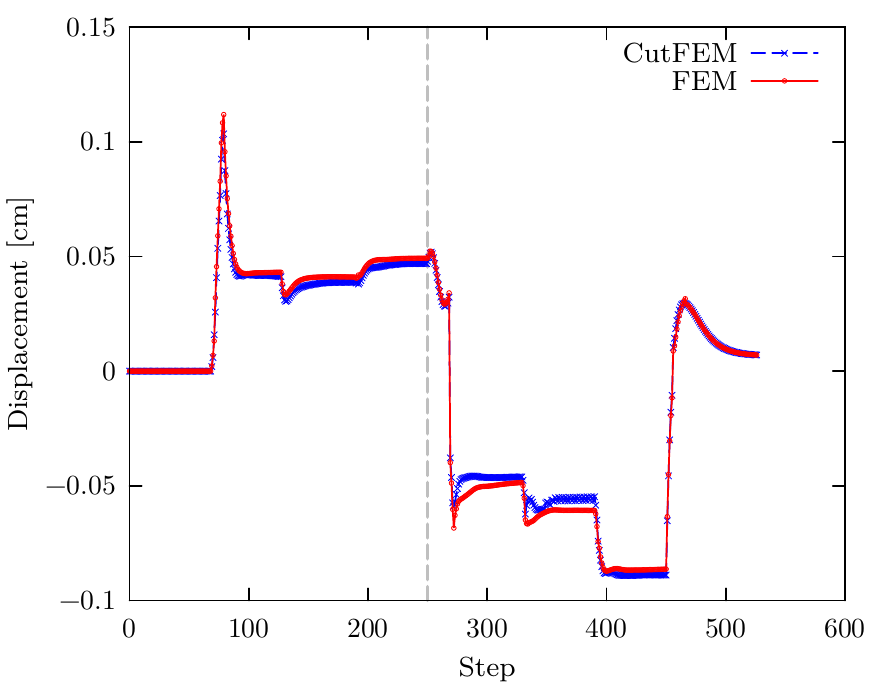}
\caption{Displacement measured at the point during insertion and retraction, using CutFEM and standard FEM. The vertical dash line indicates the moment when the needle is starting to be retracted.}
\label{fig:needle_liver/displacement}
\end{figure}

As can be seen in \cref{fig:needle_liver/displacement} that, the displacement at the point during insertion and retraction, simulated with CutFEM agrees well with that simulated by standard FEM.

\subsection{Electrode implantation simulation in Deep Brain Stimulation}




The cut FEM is now employed to simulate an electrode lead implantation, using in Deep Brain Stimulation (DBS) procedure. We also take into account the brain shift phenomenon due to the leak of cerebro-spinal fluid when a burr hole is drilled in the skull to access the brain tissue. The goal of the simulation is to insert an electrode inside the brain until it reaches the subthalamic nucleus (STN) area for treatment of Parkinson's disease. To do that, a cannula is inserted together with the electrode lead through a hole drilled in the skull. When they reach the STN area, the cannula is retracted while keeping the electrode lead inside. As in \cite{Bui_2017_error_brain}, frictional interactions between the brain tissue with the cannula and electrode lead are simulated.

Young's modulus of $6$ kPa and Poisson's ratio of $0.45$ are set to the brain tissue. The cannula and electrode lead are set with Young's modulus of $10$ GPa, and with Poisson's ratio of $0.3$. 

\begin{figure}[!htbp]
 \centering
       \begin{subfigure}[b]{0.5\textwidth}
	  \centering
	  \includegraphics[width=1\columnwidth]{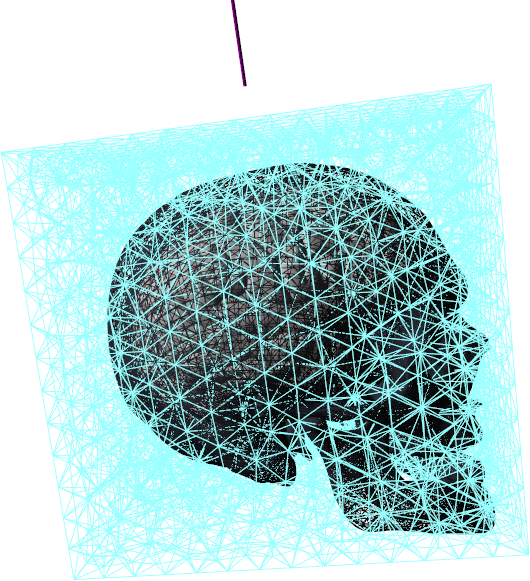}
	  \caption{}
	  \label{fig:brain_cannula_backgroundMesh}
      \end{subfigure}%
      	~ 
      \begin{subfigure}[b]{0.5\textwidth}
              \centering
	      \includegraphics[width=0.72\columnwidth]{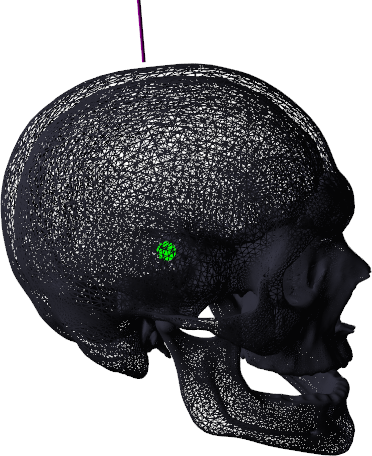}
              \caption{}
              \label{fig:STN_target_position}
      \end{subfigure}%
\caption{Input background mesh used for simulation of brain behaviour (\subref{fig:brain_cannula_backgroundMesh}), and the displacement is measured, during simulation, at the point shown in green colour in (\subref{fig:STN_target_position}).}\label{fig:brain_shift_input}
\end{figure}

The input background mesh used for the cut FEM simulation of brain behaviour is shown in \cref{fig:brain_cannula_backgroundMesh}. We consider simple boundary conditions for the brain tissue. Indeed, brain tissue around the optic nerves and the brainstem are considered to be clamped. And, bilateral interaction constraints are considered between the brain surface and the skull. It is noted that these constraints are implicitly applied on the brain surface as described in \cref{sec:boundary_condition_on_immersed_surfaces}. During simulation, displacement is measured at the point in the STN area as shown in \cref{fig:STN_target_position}.

\begin{figure}[!htbp]
 \centering
\includegraphics[width=1\columnwidth]{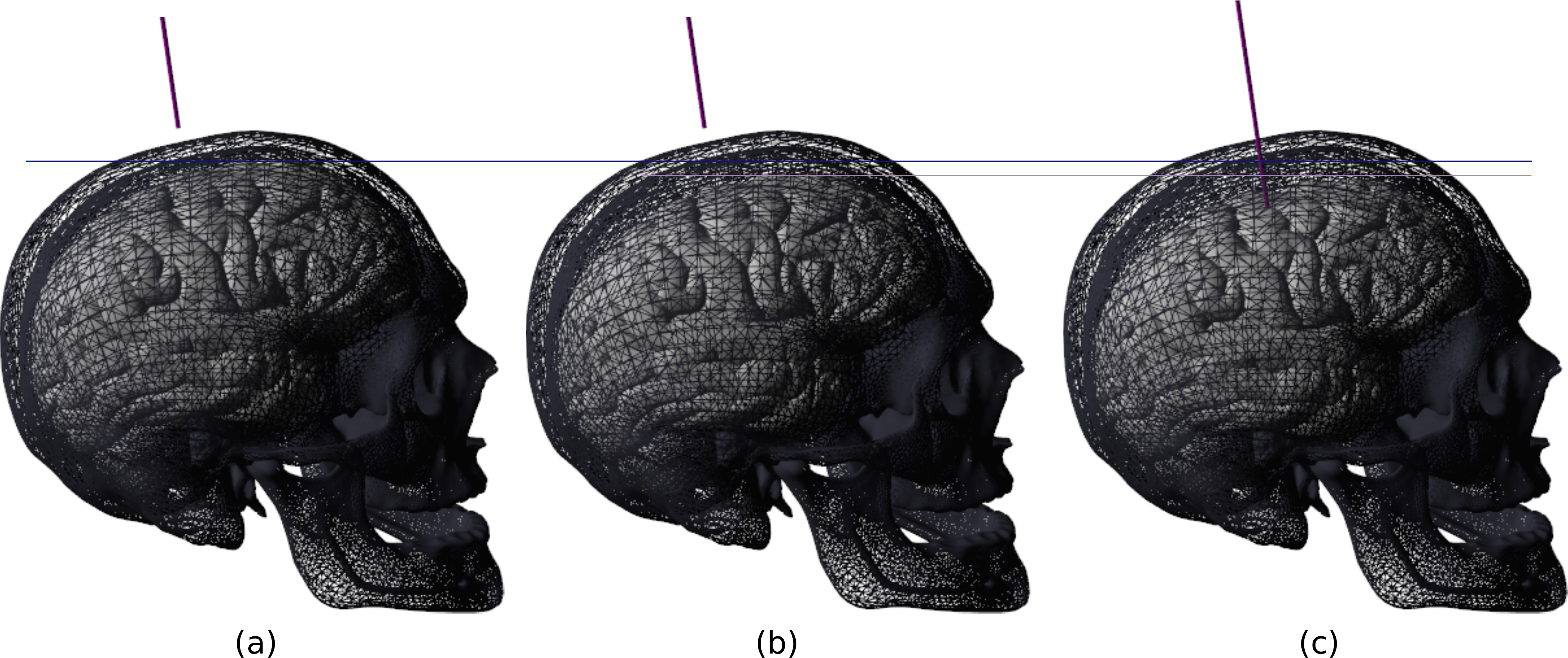}
\caption{Brain at initial state (a), and brain deformation due to brain shift (b), and brain deformation due to cannula insertion (c). The horizontal lines help to visualise the brain deformations.}
\label{fig:brain_init_shifted_deformedByCannula}
\end{figure}

\begin{figure}[!htbp]
 \centering
\includegraphics[width=0.75\columnwidth]{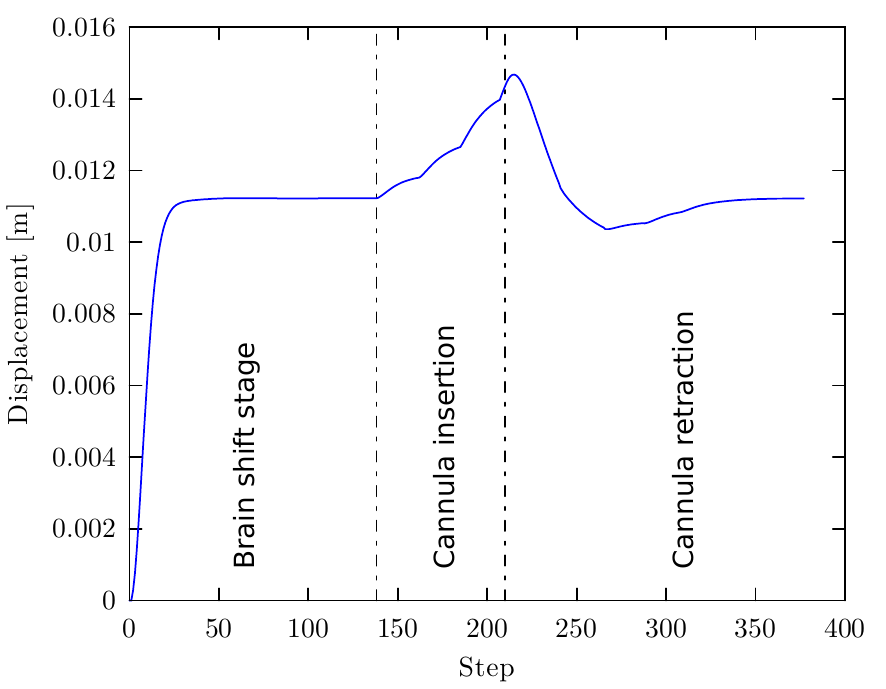}
\caption{Displacement of the STN target due to brain shift, and due to cannula insertion and retraction.}
\label{fig:brain/displacement_parkinson_target}
\end{figure}

\cref{fig:brain_init_shifted_deformedByCannula} shows the brain deformation at different stages of the simulation. The horizontal lines in \cref{fig:brain_init_shifted_deformedByCannula} help to show the differences of the brain deformation due to brain shift, and due to cannula insertion. The displacement of the STN target due to brain shift, and due to cannula insertion and retraction, is presented in \cref{fig:brain/displacement_parkinson_target}. It reveals that the brain shift is the origin of the STN displacement which is stabilised about $1.1$~cm. When the cannula is inserted inside the brain tissue, due to frictional interaction between them, the displacement of the STN target increases. And just after the cannula tip has reached the STN target, the cannula is undergoing retraction, and this induces the decreasing of the STN displacement before the STN target is stabilised around the location, when the cannula is fully retracted from the brain tissue, where the STN was found after the brain shift stage.

These results prove that the cut FEM is able to physically simulate the behaviour of the brain while it is in interactions with the skull, and with the cannula and electrode lead, even a nonconforming mesh is used. This makes the discretisation as independent as possible from the geometric description. We believe that, with such a tool, patient-specific simulations (in term of geometries for instant) can be performed, particularly in the context of real-time simulations.

\section{Conclusions}
\label{sec:Conclusions}

A corotational formulation of cut finite element method has been proposed. We have also shown the methods, using Lagrange multipliers, to apply Dirichlet boundary conditions implicitly on the immersed surface which is not conformed to the background mesh used for simulations. We verified the implementation by studying convergences through a tensile test, and a bending test as well. We also demonstrated the performance of cut FEM, compared to standard FEM, through needle insertion problems in medical simulations. By employing the cut FEM, it makes the discretisation as independent as possible from the geometric description, and also minimises the complexity of mesh generation, especially for complex geometries. Using coarse meshes, because of the constraint of computational time in real-time simulation context, while still preserving the geometric details is becoming possible by using cut FEM.

Two kinds of applications using the cut FEM have been studied in this paper: 
\begin{inparaenum}[i)]
  \item immersed interfaces, and
  \item fictitious boundaries.                                             
\end{inparaenum}
The immersed interfaces are useful for simulations of heterogeneity of tissues, \eg{}, when tumours or internal structures of tissues are considered. Simulations using fictitious boundaries are suitable for applications in which the tissue geometries are complex, and by using the cut FEM, it is possible to integrate only the material inside the tissue surface while using a nonconforming mesh to its boundaries for the simulation.

Interaction between surgical tools, \eg{}, the needle, with the tissue surface being implicitly defined, has also been shown to be working properly.

One limitation of our work is that we use linear elements for discretisation of simulated domains, and also for embedding of subelements for cut elements to facilitate numerical integration. By doing this, we still cannot precisely capture the surface geometries. The next step should be using higher order elements for more accurate integration of implicit geometries, as proposed in \cite{Fries2016}.

\section*{Acknowledgements}
\label{sec:Acknowledgements}

St\'ephane Bordas, Satyendra Tomar and Huu Phuoc Bui thank partial funding for their time provided by the European Research Council Starting Independent Research Grant (ERC Stg grant agreement No. 279578) RealTCut ``Towards real time multiscale simulation of cutting in non-linear materials with applications to surgical simulation and computer guided surgery''. We also also grateful for the funding from the Luxembourg National Research Fund (INTER/MOBILITY/14/8813215/CBM/Bordas and INTER/FWO/15/10318764).

\bibliographystyle{wileyj}
\bibliography{refs}

\end{document}